\documentclass[review,3p,11pt]{elsarticle}

\usepackage{amssymb}
\usepackage{amsthm}

\usepackage{amsmath}
\usepackage{bm}
\usepackage{subfigure}
\usepackage{graphics}
\usepackage{here}
\usepackage{color}
\usepackage{ulem}

\def\XXint#1#2#3{{\setbox0=\hbox{$#1{#2#3}{\int}$ }
		\vcenter{\hbox{$#2#3$ }}\kern-.6\wd0}}

\definecolor{mygray}{gray}{0.5}

\makeatother

\journal{International Journal of Heat and Mass Transfer}
\begin{document}

\begin{frontmatter}



\title{
Level set-based multiscale topology optimization\\ for a thermal cloak design problem using the homogenization method
}


%
\address[UT]{Department of Mechanical Engineering, Graduate School of Engineering, The University of Tokyo, Yayoi 2-11-16, Bunkyo--ku, Tokyo 113-8656, Japan.}
\address[UT2]{Department of Strategic Studies, Institute of Engineering Innovation, School of Engineering, The University of Tokyo, Yayoi 2-11-16, Bunkyo-ku, Tokyo 113-8656, Japan}
\author[UT]{Makoto Nakagawa}
\author[UT,UT2]{Yuki Noguchi}
\author[UT]{Kei Matsushima}
\author[UT,UT2]{Takayuki Yamada\corref{cor1}}
\ead{t.yamada@mech.t.u-tokyo.ac.jp}
\cortext[cor1]{Corresponding author.
	Tel.: +81-3-5841-0294;
	Fax: +81-3-5841-0294.}

\begin{abstract}
%

\textcolor{black}{Artificially designed composite materials consist of microstructures, that exhibit various thermal properties depending on their shapes, such as anisotropic thermal conductivity.}
One of the representative applications of such \textcolor{black}{composite materials for thermal control} is the thermal cloak. This study proposed a topology optimization method based on a level set method for a heat conduction problem to optimally design \textcolor{black}{composite materials} that achieve a thermal cloak. The homogenization method was introduced to evaluate \textcolor{black}{its} effective thermal conductivity coefficient. 
Then, we formulated a multiscale topology optimization method for \textcolor{black}{the composite materials} in the framework of the homogenization method, where the microstructures were optimized to minimize objective functions defined using the macroscopic temperature field. 
We presented examples of optimal structures in a two-dimensional problem and discussed the validity of the obtained structures.
\end{abstract}


\end{frontmatter}

\section{Introduction}\label{sec:introduction}
%
Heat flux control in thermal conduction is important in various applications, such as 
energy management of batteries~\cite{haoEfficientThermalManagement2018}, 
thermal dissipation in high-density integrated circuits~\cite{prasadReviewSelfHeatingEffects2019}, and 
reduction of thermal interfacial resistance by carbon nanotubes~\cite{kaurEnhancedThermalTransport2014}.
Focusing on integrated circuit boards, temperature-sensitive devices must be shielded from the heat generated by heat-generating devices~\cite{dedeThermalMetamaterialsHeat2018}.
In this regard, thermal cloaks efficiently control heat flux to make the temperature field around an object behave as if it were not there.
So far, various strategies have been proposed to achieve thermal cloak, such as bilayer thermal cloaks~\cite{hanExperimentalDemonstrationBilayer2014,xuUltrathinThreeDimensionalThermal2014,maExperimentalDemonstrationMultiphysics2014} \textcolor{black}{, the application of coordinate transformation~\cite{pendryControllingElectromagneticFields2006,ji2019Nonsingular},} the use of thermoelectric devices to control heat flux~\cite{nguyenActiveThermalCloak2015}, and, recently, machine learning for designing multilayer-based thermal cloaks~\cite{jiDesignThermalCloaks2022}. \textcolor{black}{Furthermore, a thermal design for cloaking has been studied on the macroscopic scale and on the nanoscale, where Fourier's law is invalid~\cite{xiao2021Inverse}.}
Among \textcolor{black}{such} strategies, the use of \textcolor{black}{thermal characteristics of artificially designed composite materials has been applied in many studies}~\cite{fanShapedGradedMaterials2008,shaRobustlyPrintableFreeform2021,sunDesignThermalCloak2022a}.

\textcolor{black}{Composite materials consist of many fine unit-cell structures arranged in periodic patterns.} 
\textcolor{black}{Among them, metamaterials have attracted researcher's interests in many fields.}
One of the characteristics of metamaterials is that they exhibit physical properties that cannot be realized using only homogeneous materials found in nature. Recently, metamaterials have been widely used in various physical fields, such as ray-optics cloaks in electromagnetics~\cite{chenRayopticsCloakingDevices2013}, acoustic cloaks~\cite{zhangBroadbandAcousticCloak2011},  and metamaterials with negative Poisson's ratio in elasticity~\cite{vogiatzisTopologyOptimizationMultimaterial2017}. 
\textcolor{black}{The concept of these metamaterials has been extended to heat conduction problems, such as} \textcolor{black}{metamaterials whose effective thermal conductivity depends on temperature \cite{liTemperatureDependentTransformationThermotics2015} and those with apparently negative thermal conductivity \cite{fanShapedGradedMaterials2008}.}
\textcolor{black}{However, although innovative heat manipulation is realized by the fruitful thermal properties of artificially designed composite materials, }
the design degree of freedom and the shape complexity \textcolor{black}{of composite materials} make it difficult for designers to realize the structural design of \textcolor{black}{composite material} with desired properties on the basis of the trial-and-error approach.

Topology optimization, initially proposed by Bends{\o}e and Kikuchi~\cite{bendsoeGeneratingOptimalTopologies1988}, is one of the solutions to overcome the above issue.
Among different structural optimization methods, the topology optimization method has the most flexible degree of design freedom based on the replacement of a structural optimization problem with a material distribution problem.
Topology optimization finds the material distribution to minimize an objective function, which is defined by the desired property of a structure. The method has been applied to a wide range of optimal design problems, such as elasticity~\cite{deleonStressconstrainedTopologyOptimization2015,emmendoerferjr.LevelSetTopology2018,emmendoerferTopologyOptimizationLocal2016},  acoustics~\cite{isakariTopologyOptimisationThreedimensional2014,lanznasterLevelsetApproachBased2021,luTopologyOptimizationAcoustic2013},  fluids~\cite{yajiTopologyOptimizationUsing2014,yuThreedimensionalTopologyOptimization2020,sasakiTopologyOptimizationFluid2019}, and thermal problems~\cite{dboukReviewEngineeringDesign2017, yamadaLevelSetBasedTopology2011}. 

Several studies have employed topology optimization to obtain optimized structures of heat-flux-controlling devices. For example, topology optimization has been applied for realizing 
thermal cloaking devices~\cite{fujiiExploringOptimalTopology2018,fujiiOptimizingStructuralTopology2019}, thermal undetectable concentrators that achieve simultaneously thermal cloaking and concentrating of heat flux~\cite{fujiiCloakingConcentratorThermal2020a,hirasawaExperimentalDemonstrationThermal2022}, thermal carpet cloaks that reduce the temperature disturbance caused by bumps~\cite{fujiiTopologyoptimizedThermalCarpet2019}, and heat flux inverters that can rotate heat flux~\cite{fachinottiOptimizationbasedDesignEasytomake2018a}.
In these examples, the macroscopic material distribution is set as a design variable; 
in other words, only a single-scale optimization is performed, and optimized structural designs are obtained on the basis of the isotropic thermal conductivity properties of the constituent materials. However, if the microstructures of the constituent materials are targeted for optimization, the design degree of freedom becomes larger, and the performance of the thermal cloak is expected to be further improved by the anisotropic properties of the optimized microstructures.

Several studies have dealt with the optimal design of thermal cloaks by optimizing the structural design of the microstructures of \textcolor{black}{composite materials.} 
For instance, Álvarez and-Fachinotti~\cite{lvarezhostosMetamaterialElastostaticCloaking2019a} optimized the orientation angle of the microstructure of a \textcolor{black}{composite material} composed of a simple stripe shape. 
In the study of Dede and Zhou~\cite{dedeThermalMetamaterialsHeat2018}, the copper wires on a printed circuit board were set as the design variables to control heat flux, and they achieved not only a thermal cloak, but also devices that rotate and concentrate heat flux.
However, these studies only fixed the basic shape of microstructures, and further performance improvements can be expected by setting the microstructures as design variables in the framework of topology optimization.

Recently, Seo and Park~\cite{seoHeatFluxManipulation2020a} proposed a multiscale topology optimization, where the microstructures of \textcolor{black}{a composite material} were optimized to manipulate macroscopic heat flux.
They used the homogenization method to evaluate the macroscopic thermal property in the system of \textcolor{black}{the composite material.}
The homogenization method~\cite{papanicolauAsymptoticAnalysisPeriodic1978,bakhvalovHomogenisationAveragingProcesses1989} is based on the asymptotic expansion of a solution, whereby a periodic microscopic unit cell structure can be regarded as a homogeneous material whose material property is characterized by a homogenized coefficient. 
Thanks to the high efficiency of homogenization in obtaining macroscopic thermal behaviors, these authors efficiently obtained optimized designs of \textcolor{black}{the composite material to control the heat flux.}
However, they did not target an optimized design for a thermal cloak.

In this study, we propose a multiscale topology optimization method for the design of a thermal cloak based on the homogenization method. 
Specifically, several kinds of microstructures are set as design variables. 
By combining this concept with the homogenization method, various thermal conductivity anisotropies can be realized, thus increasing the degree of flexibility in design.
Two objective functions are defined to optimally design the thermal cloak by taking full advantage of the design freedom in the proposed multiscale topology optimization method: 
One is to make the temperature field outside the thermal cloak behave as if there were no object, and the other is to reduce the incoming magnitude of the heat flux, which makes any objects within the thermal cloak thermally invisible.

This paper is organized as follows. The homogenization method is briefly introduced in Section~\ref{sec: homogenization}. The problem setting for the thermal cloak is explained in Section~\ref{sec: design settings}, and a thermal cloak design problem is formulated in Section~\ref{sec:formulation}. In Section~\ref{subsec:sensitivity analysis}, sensitivity analysis is performed according to the concept of the topological derivative.
We then introduce a level set-based topology optimization method in Section~\ref{sec:level set-based topology} to solve the multiscale optimization problem. In Section~\ref{sec: numerical implementation}, we describe the specific numerical implementation in the optimization calculation. In Section~\ref{sec: numerical examples}, we show the two-dimensional optimization results to demonstrate the effectiveness of the proposed method.  
In Section~\ref{subsec:The validity of the optimization }, we show the validity of the optimized unit-cell structure by investigating the macroscopic temperature response of a \textcolor{black}{designed composite material} composed of finite numbers of the optimized unit cells.

\section{\textcolor{black}{Homogenization method for composite materials in a heat conduction problem}}\label{sec: homogenization}
\begin{figure}[H]
	\includegraphics[scale=0.7]{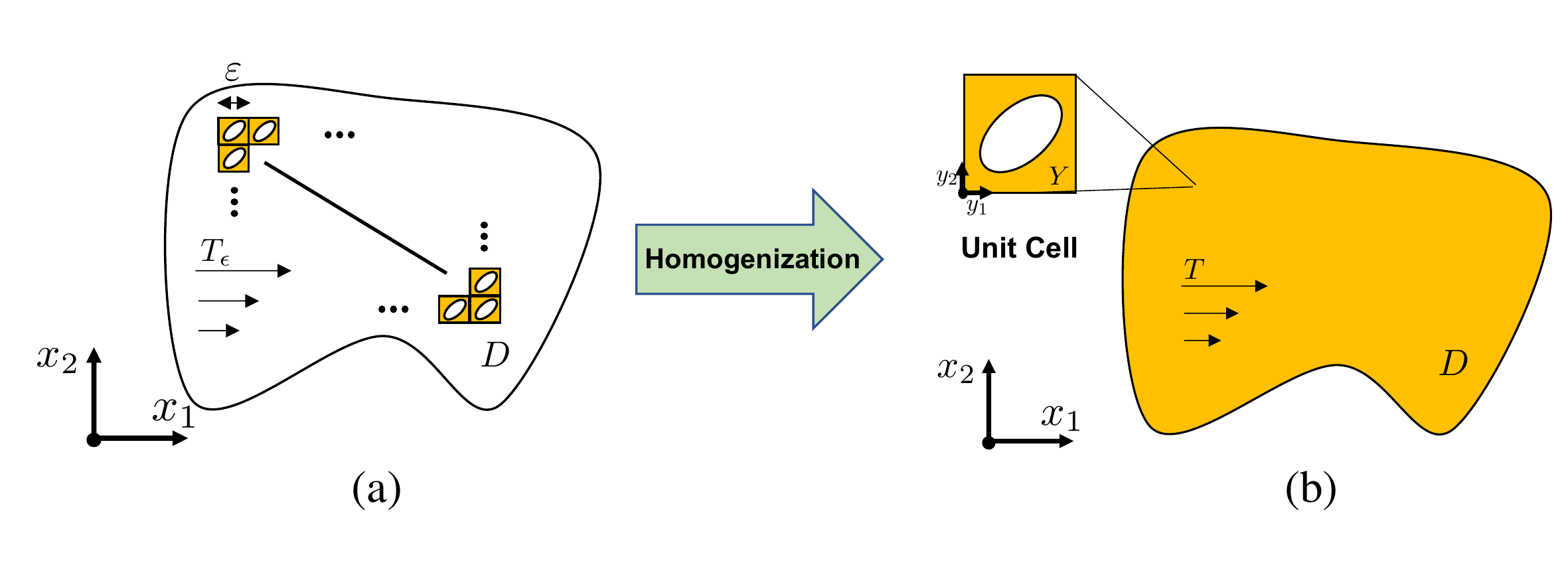}
	\caption{Concept of the homogenization method. (a) Heterogeneous media with a periodic unit-cell structure.   (b) Homogenized media replaced by the homogenization method.}
	\label{fig: homogenization}
\end{figure}
In this section, we briefly introduce the homogenization method for evaluating the \textcolor{black}{thermal properties of composite materials.}
We consider the steady-state heat conduction problem for a periodic heterogeneous media with a series of microscopic unit-cell structures of characteristic length $\epsilon$, as shown in Fig.~\ref{fig: homogenization}(a). The unit cell is composed of two homogeneous, isotropic, linear thermal conductivity materials.
On a region $D$ lined with periodic microstructures, the temperature field $T_\epsilon$ when a heat source $f$ is generated is given as the solution of a governing equation based on Fourier's law:
\begin{align}
	\begin{cases}
		-\mathrm{div} \left(K(\frac{x}{\epsilon})\nabla T_{\epsilon} \right)=f \quad &\text{in} \quad D, \\
		T_\epsilon =0 \quad &\text{on} \quad \partial D, \\
	\end{cases}
\label{eq:governing equation1}
\end{align}
where $x$ is the coordinate system defined in the macroscopic system representing the entire region and, $y=\frac{x}{\epsilon}$ is the coordinate system defined in the microstructure, which is normalized by the unit cell length $\epsilon$. 
$K(\frac{x}{\epsilon})$ is the heat conductivity coefficient, and it shows a periodic distribution corresponding to the periodic heterogeneous media. That is, $K(y)$ with $y \in Y=(0,1)^N$ satisfies $K(y)=K(y+e_i)$, where $N=2$ or $3$ is the spatial dimension, $Y$ represents the unit cell, and $e_i$ ($1\le i \le N$) represents the orthogonal basis in the microscale coordinate. 
\textcolor{black}{The temperature field $T_\epsilon$ is the difference between a reference temperature (e.g., room temperature) and the thermodynamic temperature.}
We assume that $T_\epsilon$ can be asymptotically expanded when the size of unit cell $\epsilon$ approaches zero as follows:
\begin{align}
	T_\epsilon(x)=\sum_{i=0}^{+\infty}\epsilon^{i}T_i\left(x,y\right),
	\label{eq:Tepsilon}
\end{align}
where $T_i(x,y)$ is an asymptotically expanded temperature field depending on both the macroscale $x$ and microscale $y$. Furthermore, using the chain rule under the assumption of Eq.~(\ref{eq:Tepsilon}), we can derive the following equation:
\begin{align}
	\nabla T_\epsilon (x)=\epsilon^{-1} \nabla_y T_0 \left(x,\frac{x}{\epsilon} \right) +\sum_{i=0}^{+\infty}\epsilon^i(\nabla_y T_{i+1}+\nabla_x T_i)\left(x,\frac{x}{\epsilon}\right).
	\label{eq:Tasym}
\end{align}
Substituting these into Eq.~(\ref{eq:governing equation1}) and comparing the terms with respect to  $\epsilon$, we obtain the following boundary value problem, called the homogenized problem:
		\begin{align}
			\begin{cases}
				-\mathrm{div}_x \left(K^*\nabla_x T \right)=f \quad &\text{in} \quad D,\\
				T =0 \quad &\text{on} \quad \partial D, \\
			\end{cases}
				\label{eq:gov}
		\end{align}
	Eq.~(\ref{eq:gov}) represents the homogeneous linear steady-state heat conduction problem with the second-order tensor $K^{*}$.
This tensor characterizes the macroscopic thermal conductivity of the microstructure and is called the homogenized coefficient or homogenized thermal conductivity tensor, given by the following equation:
\begin{align}
	K_{ij}^{*}=\int_{Y} K(y)(e_i+\nabla_y w_i)\cdot (e_j+\nabla_y w_j) dy \quad {(i, j=1, ..., N)}.
	\label{eq:homogezied coefficient}
\end{align}
$w_{i,j}$ is obtained by solving the unit cell problem Eq.~(\ref{eq:cell problem}) as follows:
\begin{align}
	\begin{cases}
		-\mathrm{div}_y \left(K(y)(e_i+\nabla_y w_i(y))\right)=0 \quad &\text{in} \quad Y,\\
		y \rightarrow w_i(y) & \quad Y\text{-periodic}.
		\label{eq:cell problem}
	\end{cases}
\end{align}

We first solve the unit cell problem in Eq.~(\ref{eq:cell problem}) to obtain the macroscopic temperature field $T$. Then, the homogenized coefficient was evaluated using Eq.~(\ref{eq:homogezied coefficient}). Finally, the homogenized problem in Eq.~(\ref{eq:gov}) is solved using the evaluated homogenized coefficient. 
\section{Topology optimization}\label{sec: topology optimization}
\subsection{Problem Settings and Governing Equation}\label{sec: design settings}
\begin{figure}[H]
	\centering
	\includegraphics[scale=0.8]{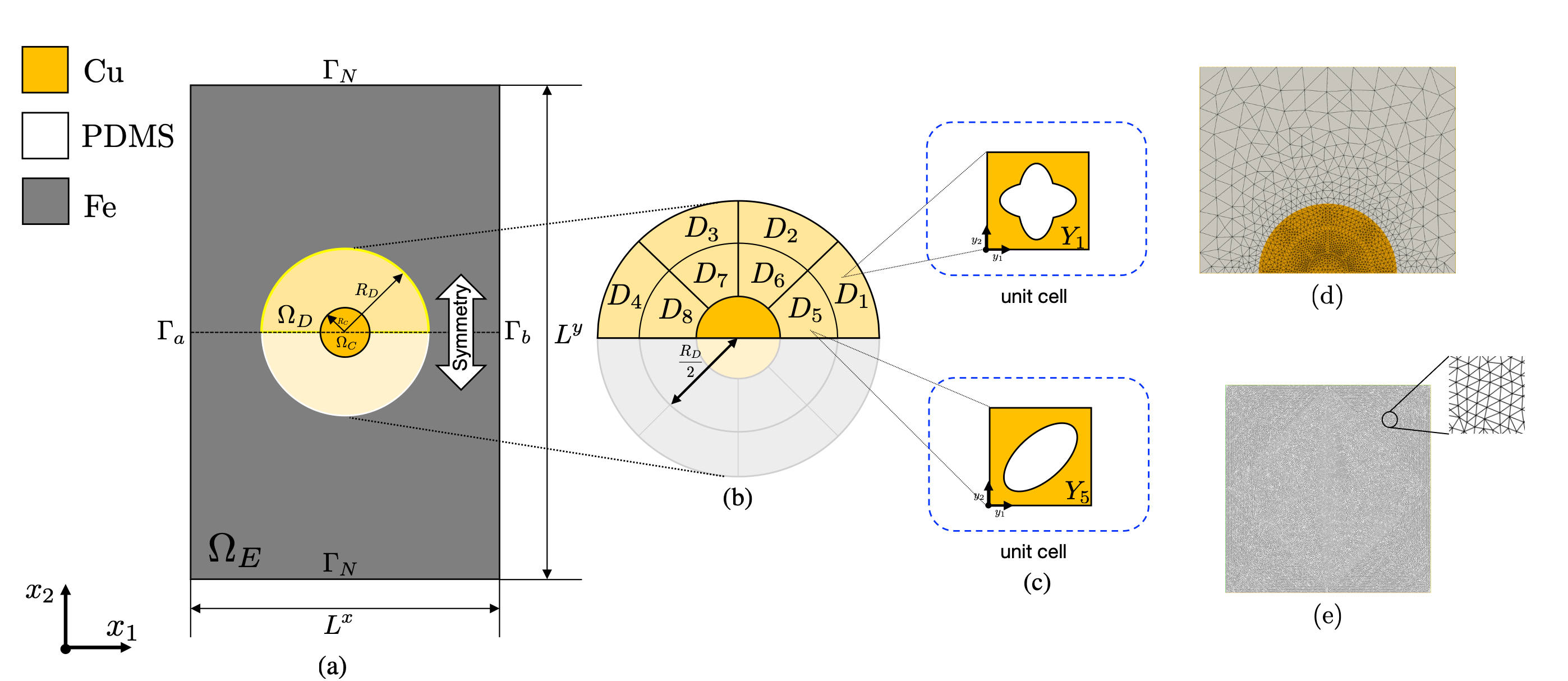}
	\caption{\textcolor{black}{(a) Geometry settings of the thermal cloak design problem.  $\Omega_E$ is the evaluation domain, $\Omega_D$ is the domain containing the optimized microstructures, and $\Omega_C$ is a domain filled with copper. The characteristic lengths of this structure are $R_D=1.35\ \text{m}$,  $L^x=5\ \text{m}$,  $L^y=8\ \text{m}$, and  $R_c=0.4\ \text{m}$.   (b) The magnified view around $\Omega_D$ divided into eight regions, $D_l$ with $1 \le l \le 8$. (c) A unit cell $Y_l$ with $ 1\le l \le 8$ with infinitesimal size is periodically laid out in $D_l$. (d) Finite element discretization of the computational domain in the macroscale. (e) Finite element discretization of the unit cell.}}
	\label{fig:Design_model} 
\end{figure}
In this study, we aim to realize two kinds of thermal properties using topology optimization. One is the ``thermal cloak", which brings the temperature field of $\Omega_E$ closer to a certain reference temperature field. We focus on the case where this reference temperature field is for an entire domain made of steel; thus, the reference temperature field is denoted by $T_{\text{steel}}$. 
The other is the ``no heat flux", which prevents heat flux in $\Omega_C$.
This allows us to achieve a thermal cloak in $\Omega_E$, no matter what object is placed in $\Omega_C$.

 The problem setting of the topology optimization for the thermal cloak is presented in Fig.~\ref{fig:Design_model}. 
\textcolor{black}{For brevity, we describe the problem in a two-dimensional setting.}
In the macroscale, the computational domain is composed of $\Omega_E$, $\Omega_D$, and $\Omega_C$ in Fig.~\ref{fig:Design_model}(a).
$\Omega_E$ is the evaluation domain for the thermal cloak filled with steel.
$\Omega_D$ is the domain where the microstructures are placed, and it is divided into several design domains $D_l$ with $1\le l \le 8$ as shown in Fig.~\ref{fig:Design_model}(b).
As shown in Fig.~\ref{fig:Design_model}(c), a microstructure made of copper ($Y_l^c$) and PDMS ($Y_l^p$) defined in the unit cell $Y_l$ is assigned to the domain $D_l$ with $1\le l \le 8$. 
$\Omega_C$ is an obstacle for thermal cloaking and is assumed to be made of copper. 
For each of the design domains $D_l$ with $1\le l \le 8$, the thermal conductivity tensor is determined using the homogenization method, as explained in Section \ref{sec: homogenization} to obtain the macroscopic temperature field for the entire domain.
The boundary condition on $\Gamma_N$ is the adiabatic condition, whereas those on $\Gamma_a$ and $\Gamma_b$ are the temperatures fixed to certain values (low temperature $T_\text{low}$ and high temperature $T_\text{high}$, respectively). 
Thus, the boundary value problem for the system shown in Fig.~\ref{fig:Design_model} is summarized as follows:
\begin{align}
	\begin{cases}
	-\mathrm{div}_{x}\left(K(\bm{x})\nabla_x T \right)=0\qquad &\text{in}\qquad \Omega,	\\		
	\bm{n}\cdot\left(K(\bm{x})\nabla_x T\right) =0 \qquad &\text{on}\qquad \Gamma_N, \\
	T =T_\text{low} \quad &\text{on} \qquad \Gamma_a, \\
	T =T_\text{high} \quad &\text{on} \qquad \Gamma_b,
	\end{cases}
\label{eq: g.v.e.q}
\end{align}
where $\bm{n} $ is the unit normal vector on $\Gamma_N$, and $K$ is the thermal conductivity tensor given as follows:
\begin{align}
	K(\bm{x})=
	\begin{cases}
		\begin{bmatrix}
			K_{l11}^{*} \quad K_{l12}^{*} \\
			K_{l21}^{*} \quad K_{l22}^{*}		
		\end{bmatrix}\quad & \text{if~~~} \bm{x} \in \quad D_{l}\ (1\le l\le 8),\\
		K_\text{steel}  \bm{I}& \text{if~~~} \bm{x} \in \quad \Omega_E ,\\
		K_\text{copper} \bm{I}& \text{if~~~} \bm{x} \in \quad \Omega_C, 
	\end{cases}
\label{eq:thermal conduct}
\end{align}
where $\bm{I} $ is the identity matrix. The thermal conductivity tensor $K_l^{*}$ in $D_l$ with $1\le l\le8$ can be evaluated using Eq.~(\ref{eq:homogezied coefficient}) from the solution of the unit-cell problem Eq.~(\ref{eq:cell problem}) for each unit cell $Y_l$. $K_\text{steel}$ is the thermal conductivity of steel, whereas $K_\text{copper}$ is the thermal conductivity of copper.
 Because the problem setting is vertically symmetric, only the top half of the regions is considered the computational domain. 
 \textcolor{black}{Section~\ref{sec: homogenization} focuses on heat transfer based on the linear heat conduction equation, and the variable $T$ is the difference between a reference temperature and the thermodynamic temperature.
 We set $T_{\text{low}}=0~\text{K}, T_{\text{high}}=1~\text{K}$ to apply a temperature difference of $1~\text{K}$ in the macroscale system.}

\textcolor{black}{The three-dimensional cloaking design problem can also be formulated as a two-dimensional one. 
In the two-dimensional case, we decompose the macroscopic domain $\Omega_D$ into regions $D_l$ with a certain angle. 
Similarly in the three-dimensional case, we can define $D_l$ by decomposing $\Omega_D$ with a certain solid angle. 
A microstructure in the square lattice is assigned to each $D_l$ when $N=2$, whereas that in the cubic lattice is assigned to each $D_l$ when $N=3$.
}
\textcolor{black}{Fig.~\ref{fig:Design_model}(d) and (e) show our triangular mesh we used. The computational domain in the macroscale was discretized with $2464$ triangular elements, whereas the unit cell was discretized with $60,838$ elements.}

\subsection{Formulation of the optimization problem}\label{sec:formulation}
\begin{figure}[H]
			\centering
			\includegraphics[scale=0.7]{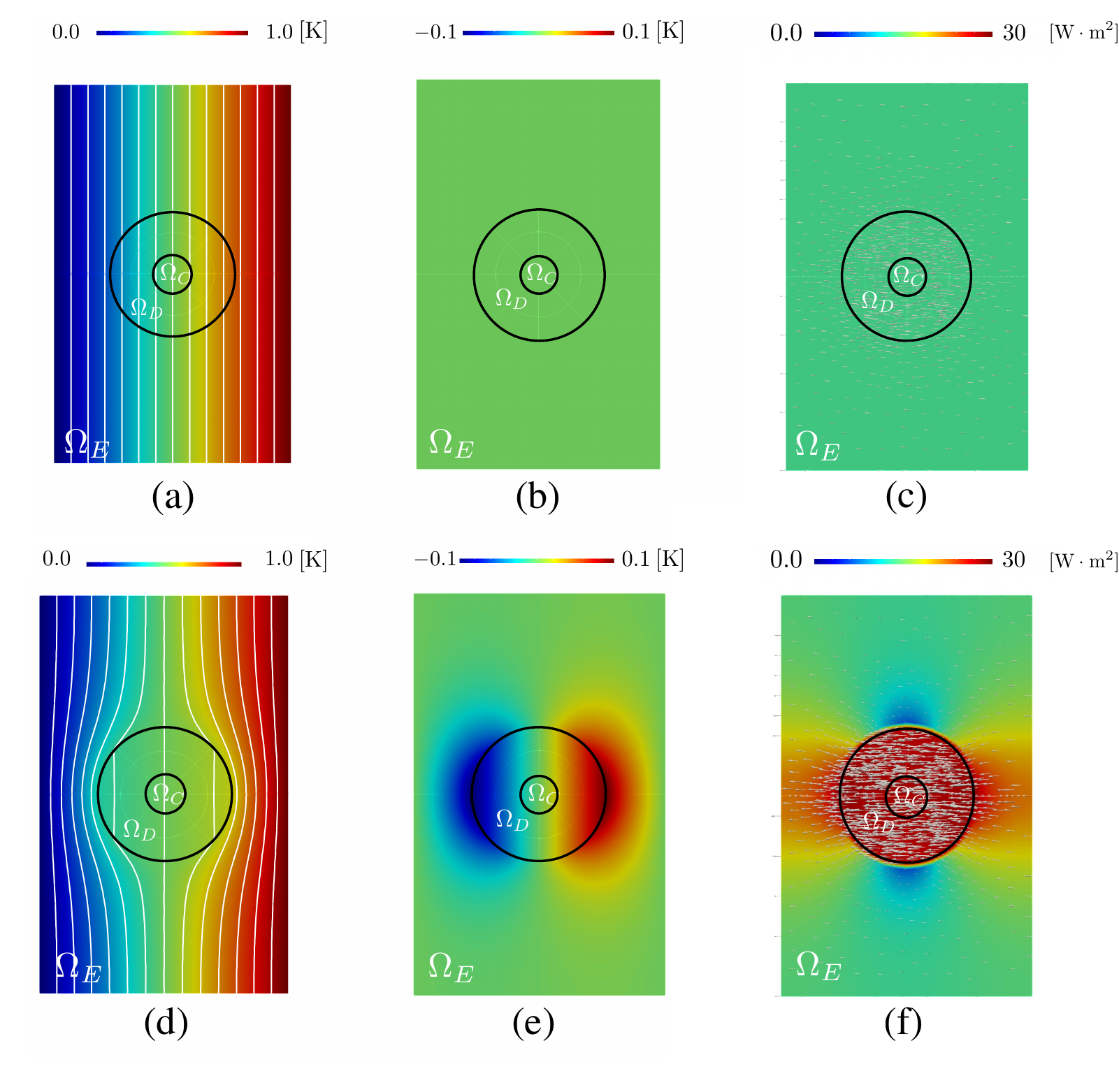}
			\caption{(a) Distribution of the reference temperature field $T_\text{steel}$ when all domains are filled with steel. (b) Distribution of $T_\text{sub}$ defined as $T_\text{sub}=T-T_\text{steel}$. (c) Magnitude and direction of the heat flux when $\Omega_D$ is filled with steel. (d) Distribution of the temperature field when $\Omega_D$ is filled with copper. (e) Distribution of $T_\text{sub}$ when $\Omega_D$ is filled with copper. (f) Magnitude and direction of the heat flux when $\Omega_D$ is filled with copper. }
\label{fig:reference}
\end{figure}
For designing of \textcolor{black}{composite materials} that achieve thermal cloak in $\Omega_E$ and no heat flux in $\Omega_C$, we define the objective function expressed by the macroscopic temperature field as follows:
\begin{align}
	J&=wJ_1 + (1-w)J_2,\nonumber\\
	J_1&= \int_{\Omega_E}(T-T_\text{steel})^2 d\Omega,~~~~~
	J_2=\int_{\Omega_C}\nabla T \cdot \nabla T d\Omega,
	\label{eq: objective functional}
\end{align}
where $0 \le w \le 1$ represents the weighting factor.
By minimizing $J_1$, the temperature field of $\Omega_E$ gets close to $T_\text{steel}$ which is a reference temperature field to be reproduced when $\Omega_D$ and $\Omega_C$ are filled with steel in Fig.~\ref{fig:reference}(a).  In other words, a thermal cloak is achieved by minimizing $J_1$.
Furthermore, the objective function for preventing heat flux from flowing to $\Omega_C$ is written as $J_2$.
By minimizing $J_2$, the heat flux in $\Omega_C$ can be reduced.
Therefore, by simultaneously minimizing $J_1$ and $J_2$, a thermal cloak can be achieved in $\Omega_E$ even if any objects are present in $\Omega_C$ as explained in Section~\ref{sec: design settings}.
In topology optimization based on the homogenization method, the objective function $J$ is calculated using homogenized coefficients in $D_l$ with $1\le l \le 8$ characterized by the unit cell structure in $Y_l$ with $1\le l \le 8$, and the unit cell structures are optimized to minimize the objective function.
Then, the optimization problem is formulated as follows:
\begin{align}
	\min_{Y_1^c ..., Y_8^c}~&J\nonumber\\
	\mathrm{subject~to~}&\mathrm{Governing~equations~(Eq.~(\ref{eq:cell problem}))~in~}Y_l~(1 \le l \le 8),\nonumber\\
	&\mathrm{Governing~equations~(Eq.~(\ref{eq: g.v.e.q}))~in~}\Omega_E \mathrm{~and~} \Omega_D \mathrm{~and~in~} \Omega_C,	\label{eq:Optimizationall}\\
	&\mathrm{the expressions~of~}(K_{l}^{*})~(1 \le l \le 8) ~\text{(Eq.~(\ref{eq:homogezied coefficient}))}.\nonumber
\end{align}
\subsection{Sensitivity analysis} \label{subsec:sensitivity analysis}
In this study, the sensitivity analysis was performed on the basis of the topological derivative concept. The topological derivative represents the amount of change in an objective function $F$ when an infinitesimal \textcolor{black}{circular ($N=2$) or spherical ($N=3$)} inclusion domain filled with material $b$ ($\Omega_b$) is inserted into the domain filled with material $a$ ($\Omega_a$).  Then, the topological derivative $D_T^{a \rightarrow b} F$ is defined as follows:
\begin{align}
	D_T^{a \rightarrow b} F =\lim_{\varepsilon \to 0} \frac{F(\Omega_a\setminus \overline{\Omega_b})-F(\Omega_a)}{|\Omega_b| },
\end{align}
\textcolor{black}{where $\varepsilon$ represents the radius of the inclusion domain, and $|\Omega_b|$ is a measure of the inclusion domain depending on the radius $\varepsilon$.}
 In the optimization problem in Eq.~(\ref{eq:Optimizationall}), the objective function $J_k\ (k=1,2)$ is expressed in terms of the macroscopic temperature field $T$, and the microstructure in the unit cell $Y_l$ with $1 \le l \le 8$ is the design variable. Therefore, this problem can be divided into two scales: (i) macroscopic temperature fields $T$ to the homogenized thermal conductivity tensor $K^*$ and (ii) the homogenized thermal conductivity tensor $K^*$ to the microstructure of unit cells. 
In (i), the objective function is defined using the macroscopic temperature field, and the design variable can be considered as the homogenized thermal conductivity tensor in $D_l$ with $1 \le l \le 8$, which takes continuous values.
In (ii), the objective function is defined using the homogenization coefficients and the design variables are the microstructures in each unit cell $Y_l$. Considering these factors, we finally formulate the topological derivative as follows:
\begin{align}
	D_T ^{a \rightarrow b}J_{k} &=\sum_{i=1}^N \sum_{j=1}^N \frac{\partial J_k}{\partial K^*_{lij}}D_T ^{a \rightarrow b}K^*_{lij},
	\label{eq: topological derivative_implicit}
\end{align}
where $\cfrac{\partial J_k}{\partial K_{lij}^*}\ (i,j=1, ..., N)$ represents the weighting factors used to increase or decrease the homogenized coefficients to minimize the objective function $J_k\ (k=1, 2)$. The explicit formula of  $\cfrac{\partial J_k}{\partial K^*_{lij}}$ is given as follows:
\begin{align}
	\cfrac{\partial J_k}{\partial K^*_{lij}}=-\int_{D_l}\frac{\partial T}{\partial x_i}\frac{\partial v^k_H}{\partial x_j }d\Omega.
\end{align}
$v_H^k(k=1,2)$ represents adjoint variables that depends on the objective functions $J_k\ (k=1,2)$. The details for the derivation of $\cfrac{\partial J_k}{\partial K^*_{lij}}$ and the adjoint equation for $v_H^k(k=1,2)$ is provided in \ref{sec:Appendix sensitivity analysis}.  $D_T^{a \rightarrow b}K^*_{lij}$ can be expressed using the orthogonal basis $e_{i,j}$ and the solution $w_{i,j}$ of the unit-cell problem in  Eq.~(\ref{eq:cell problem}).
\textcolor{black}{When $N=2$, it is given as the following equation}~\cite{giustiSensitivityMacroscopicThermal2008, noguchiTopologyOptimizationAcoustic2021}:
\begin{align}
	D_T^{a \rightarrow b}K^*_{lij}=\cfrac{2K_a(K_b-K_a)}{K_b+K_a}(e_i+\nabla_y w^l_i)(e_j+\nabla_y w^l_j)\quad(i,j=1,2)\quad\text{in}\quad Y_l,
	\label{eq:dtK}
\end{align}
where $K_{a, b}$ represents the thermal conductivity of material $a$, and $b$, and $w^l_{i, j} \ (i, j= 1, 2)$ is the solution of the unit-cell problem in Eq.~(\ref{eq:cell problem}) for $Y_l$. 
\textcolor{black}{In the case of $N=3$, a formula similar to Eq.~(\ref{eq:dtK}) can be obtained. However, the coefficient expressed by $K_{a, b}$ is different, which can be derived following the procedure in \cite{Amstutz2006}. }
As mentioned in Section~\ref{sec:formulation}, this study considered two objective functions. Therefore, we must set up the two topological derivatives $D_T J_{l1}$, and $D_T J_{l2}$ using, $v_H^k \ (k=1, 2)$ as follows:
\begin{align}
	D_T ^{a \rightarrow b}J_{1l} &=-\sum_{i=1}^{N}\sum_{j=1}^{N}\left(\int_{D_l}\frac{\partial T}{\partial x_i}\frac{\partial v_H^1}{\partial x_j }d\Omega\right) D_T ^{a \rightarrow b}K^*_{lij},\\
		D_T ^{a \rightarrow b}J_{2l} &=-\sum_{i=1}^{N}\sum_{j=1}^{N}\left (\int_{D_l}\frac{\partial T}{\partial x_i}\frac{\partial v_H^2}{\partial x_j }d\Omega \right)D_T ^{a \rightarrow b}K^*_{lij}.
\end{align}
\subsection{Level set-based topology optimization}\label{sec:level set-based topology}
We optimized the multiple unit-cell structures set up in Section~\ref{sec: design settings} to minimize the objective function Eq.~(\ref{eq: objective functional}). A level set-based topology optimization method~\cite{yamadaTopologyOptimizationMethod2010a} is thus introduced to solve such an optimization problem. In the level set method, the material distribution is represented by a scalar function, named the level set function. As shown in Section~\ref{sec: design settings}, the unit cell structure $Y_l$ in $D_l$ with $1 \le l \le 8$ is composed of two different materials: PDMS and copper. The material distribution in $Y_l$ is expressed in terms of a level set function $\phi_l$ with $1 \le l \le 8$, as in the following equation, with negative values if filled with PDMS ($Y_l^p$), zero value representing the structural boundaries ($\Gamma_l$), and positive values if filled with copper 
 ($Y_l^c$).
\begin{eqnarray}
	\left\{
	\begin{array}{ll}
		0<\phi_l(\bm{y})\le 1 &\mathrm{if}~~\bm{y}\in Y_l^c,\\
		\phi_l(\bm{y})= 0 &\mathrm{if}~~\bm{y}\in \Gamma_l,\\
		-1\le \phi_l(\bm{y})< 0 &\mathrm{if}~~\bm{y}\in Y_l^p,\label{eq:profile of LSF}
	\end{array}
	\right.
\end{eqnarray}
where $1\le l \le 8$. The characteristic function is defined using the level set function to represent the material distribution in each unit cell:
\begin{eqnarray}
	\chi_{\phi_l}=
	\left\{
	\begin{array}{ll}
		1 &\mathrm{if}~~\phi_l\ge 0, \\
		0 &\mathrm{if}~~\phi_l< 0.
	\end{array}
	\right.
\end{eqnarray}
Using this characteristic function, the optimization problem in Section~\ref{sec: design settings} can be replaced as
\begin{align}
	\inf_{\chi_{\phi_1}, ..., \chi_{\phi_{8}}}~~~&J.
\end{align}
Because finding 
the optimized characteristic function $\chi_{\phi_l}$ is difficult, 
the optimization problem can be replaced by the following reaction-diffusion equation with a fictitious time $t$.
\begin{align}
	\frac{\partial \phi_l}{\partial t}=-K_\phi ({J_l}'-\tau \nabla_y^2 \phi_l),
	\label{eq:reaction-diffusion}
\end{align}
where $K_\phi > 0$ is a positive-valued constant and $\tau$ is the regularization factor. ${J_l}'$ is the design sensitivity defined using the topological derivative. 
According to the definitions of the level set function and topological derivative explained in Section~\ref{subsec:sensitivity analysis}, the design sensitivity is defined using the topological derivatives $D_T J_{1l}$ and $D_T J_{2l}$. Because these functions have different dimensions, the design sensitivity $J^{'}_l$ must be normalized to avoid reflecting the influence of only one of the objective functions. Thus, two coefficients $C_1$ and $C_2$ are introduced to normalize the design sensitivity formulated as follows:
\begin{align}
	{J_l}'&=C_1(D_T J_{1l}^{p \to c}(1-\chi_{\phi_l})-D_T J_{1l}^{c \to p}\chi_{\phi_l})
	        +C_2(D_T J_{2l}^{p \to c}(1-\chi_{\phi_l})-D_T J_{2l}^{c \to p}\chi_{\phi_l}), \nonumber\\
	C_1&=\frac{w}{\int_{Y_l}| D_T J_{1l}^{p \to c}(1-\chi_{\phi_l})-D_T J_{1l}^{c \to p}\chi_{\phi_l}   |d\Omega},\\ C_2&=\frac{1-w}{\int_{Y_l}| D_T J_{2l}^{p \to c}(1-\chi_{\phi_l})-D_T J_{2l}^{c \to p}\chi_{\phi_l}   |d\Omega},\nonumber
\end{align}
where $D_T J_l^{p\to c}$ represents the change in the objective function when an infinitesimal $\Omega_c$ appears in $\Omega_p$, and $D_T J_l^{c\to p}$ is the opposite, as explained in Section~\ref{subsec:sensitivity analysis}.
\section{Numerical implementation} \label{sec: numerical implementation}
\textcolor{black}{\subsection{Approximated characteristic function}
Instead of the characteristic function in Eq.~(18), we use the following approximated characteristic function to make a numerical treatment to represent material distribution easier:
\begin{eqnarray}
	\chi_{\phi_l}=
	\left\{
	\begin{array}{ll}
		0 &\mathrm{if}~~\phi_l < -d, \\
		\frac{1}{2} + \frac{\phi}{d} (\frac{15}{16} - \frac{\phi^2}{d^2}(\frac{5}{8}-\frac{3}{16}\frac{\phi}{d^2}))&\mathrm{if}~~-d \le \phi_l < d,  \\
		1 &\mathrm{if}~~\phi_l \ge d.
	\end{array}
	\right.
\end{eqnarray}
where $d \in (0,1)$ is the transition width, typically set to a sufficiently small value. 
The detail on how to set $d$ is explained in the next subsection.
}
\subsection{Optimization algorithm}\label{sec: optimization flow}
In this study,  the optimization algorithm is based on the following steps.\\
$\bm{step\ 1:}$ Initialize the level set functions and all variables.\\ 
$\bm{step\ 2:}$ Solve the cell problems Eq.~(\ref{eq:cell problem}) to obtain $w_{i, j}$. \\
$\bm{step\ 3:}$ Evaluate the homogenization coefficients by Eq.~(\ref{eq:homogezied coefficient}) and solve the homogenized equation Eq.~(\ref{eq:gov}) to obtain the macroscopic temperature field $T$.\\
$\bm{step\ 4:}$ Compute the objective function $J_{1}, J_{2}$ and make a convergence decision. If it is not satisfied, go to $\bm{step\ 5}$.
\\
$\bm{step\ 5:}$ Compute the design sensitivities to the objective functions.\\
$\bm{step\ 6:}$ Update the level set functions for $1\le l \le 8$ using the reaction-diffusion equation Eq.~(\ref{eq:reaction-diffusion}).\\
$\bm{step\ 7:}$  Back to $\bm{step\ 2}$\\

\textcolor{black}{
We first set the transition width $d$ of the approximated characteristic function in Eq. (22) as $d=0.2$ to stabilize the optimization calculation. Then, we use $d=0.01$ to obtain the converged results after 70 iterations of the optimization calculation.
}
\textcolor{black}{
As a convergence criterion in $\bm{step\ 4}$, we refer to the value of the objective functions. Our numerical experiments showed that they could be converged with sufficient optimization iterations after fixing the value of $d$.
Therefore, we set the maximum value of the iterations and stopped the optimization calculation at the 150th iteration.
}

The finite element method was used in Steps 2-5 based on the implementation by FreeFEM \cite{hechtNewDevelopmentFreefem2012}, an open-source software of the finite element method.
For ensuring the validity of the optimized structures shown in Section~\ref{subsec:The validity of the optimization },
we used COMSOL Multiphysics, a multiphysics simulation FEM software. 

\section{Numerical examples} \label{sec: numerical examples}
\subsection{Optimization results and discussion}\label{sec: main results}

\begin{figure}[H]
			\centering
			\includegraphics[scale=0.7]{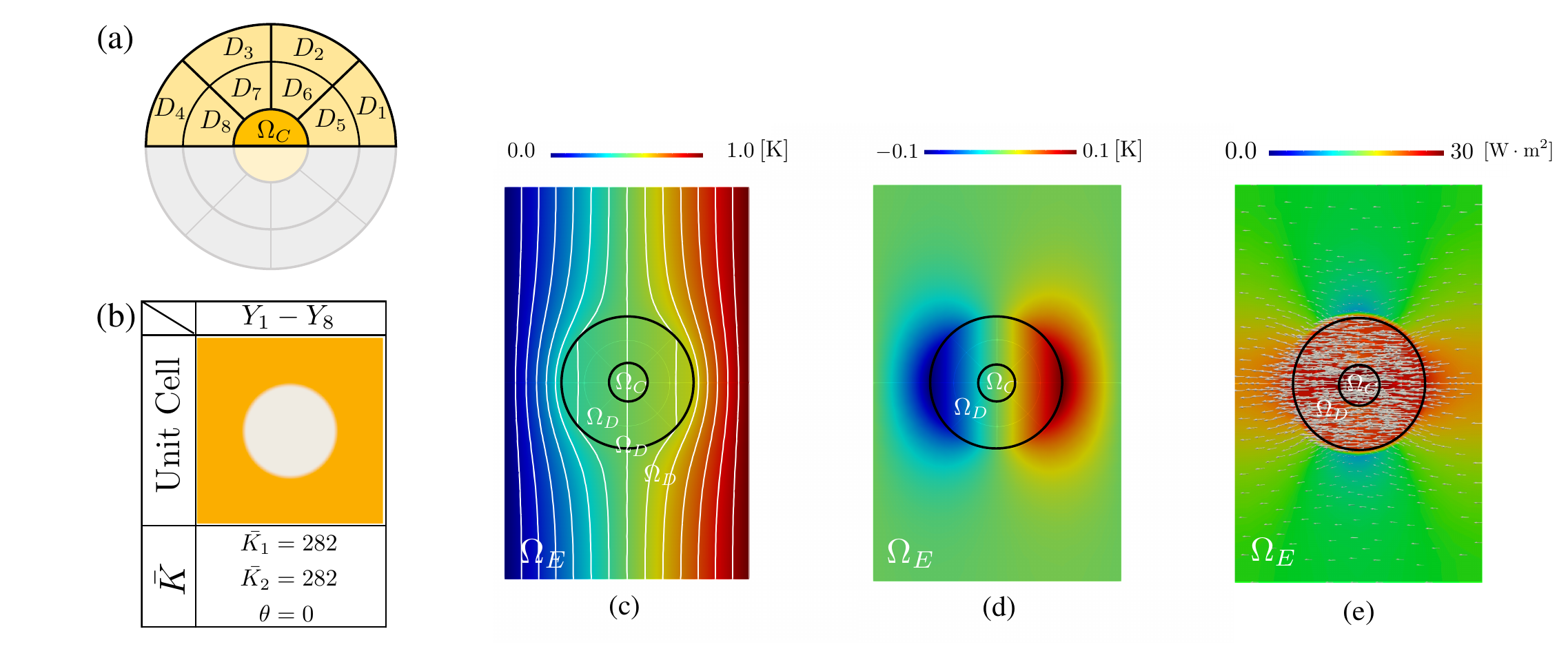}
			\caption{\textcolor{black}{(a) Arrangement of design domains $D_l$ with $1\le l \le 8$. The orange domain is filled with copper, whereas the gray domain is filled with PDMS.  (b) The initial unit cell structure for $Y_l$ with $1\le l \le 8$ and the diagonalized thermal conductivity tensor $\bar{K}$. (c) The distribution of the temperature field $T$ obtained through the homogenization. (d) The distribution of the temperature difference $T_\text{sub}$. (e) The magnitude and direction of the heat flux. }}
			\label{fig:initial struct}
\end{figure}
\textcolor{black}{In the following, we show numerical examples focusing on the two-dimensional case ($N=2$). }As explained in Section \ref{sec: design settings}, the unit cell consists of two materials: copper and PDMS. 
The arrangement of $D_l$ with $1 \le l\le 8$ and $\Omega_C$ is shown in Fig.~\ref{fig:initial struct}(a).
The initial structure of the unit cell was assumed as shown in  Fig.~\ref{fig:initial struct}(b), with PDMS of radius 0.25 of the unit-cell length placed in the center of the unit-cell. 
Fig.~\ref{fig:initial struct}(c) and (d)  show the temperature field $T$ obtained through homogenization and by the temperature difference $T_\text{sub}=T-T_\text{steel}$, respectively. The thermal conductivity of copper, PDMS, and steel were set as $K_\text{copper}=386 \ \text{W}\text{K}^{-1}\text{m}^{-1}$, $K_\text{PDMS}=0.15 \ \text{W}\text{K}^{-1}\text{m}^{-1}$, and $K_\text{steel}=67 \ \text{W}\text{K}^{-1}\text{m}^{-1}$, respectively. \textcolor{black}{The parameters in the optimization calculation, $K_{\phi}$ and $\tau$, were $K_{\phi} = 1.5$ and $\tau = 2.0\times10^{-4}$, respectively.}

The initial values of $J_1$ and $J_2$ were 
\textcolor{black}{$J_1^{\text{Init}}=2.22\times 10^{-2}\ \text{K}^2\text{m}^2$}
\textcolor{black}{and}
 \textcolor{black}{$J_2^{\text{Init}}=1.4\times 10^{-3}\ \text{K}^2$}.
 From Fig.~\ref{fig:initial struct}(d), $T_\text{sub}$ is large outside the design domain $\Omega_D$, which is not suitable for realizing the thermal cloak. Therefore, we minimized the objective function $J_1$ to reduce $T_\text{sub}$ and bring the temperature field $T$ closer to $T_\text{steel}$.

 We then introduced the diagonalized thermal conductivity tensor $\bar{K}$. The homogenized thermal conductivity tensor $K^*$ obtained through homogenization method was diagonalized to $\bar{K}$ using the rotation transform with the angle $\theta\ ^\circ$ as follows:
\begin{align}
	\bar{K_l}=
	\begin{bmatrix}
		\bar{K_{l1}} \quad 0 \\
		0 \quad \bar{K_{l2}}    
	\end{bmatrix}=
	\begin{bmatrix}
		\cos\theta \quad \sin\theta \\
		-\sin\theta \quad \cos\theta    
	\end{bmatrix}
	\begin{bmatrix}
		K_{l11}^{*} \quad K_{l12}^{*} \\
		K_{l21}^{*} \quad K_{l22}^{*}		
	\end{bmatrix}
	\begin{bmatrix}
		\cos\theta \quad -\sin\theta \\
		\sin\theta \quad \cos\theta    
	\end{bmatrix}
	\label{eq:diagonalization}
\end{align}
\begin{figure}[H]
	\centering
	\includegraphics[scale=0.5]{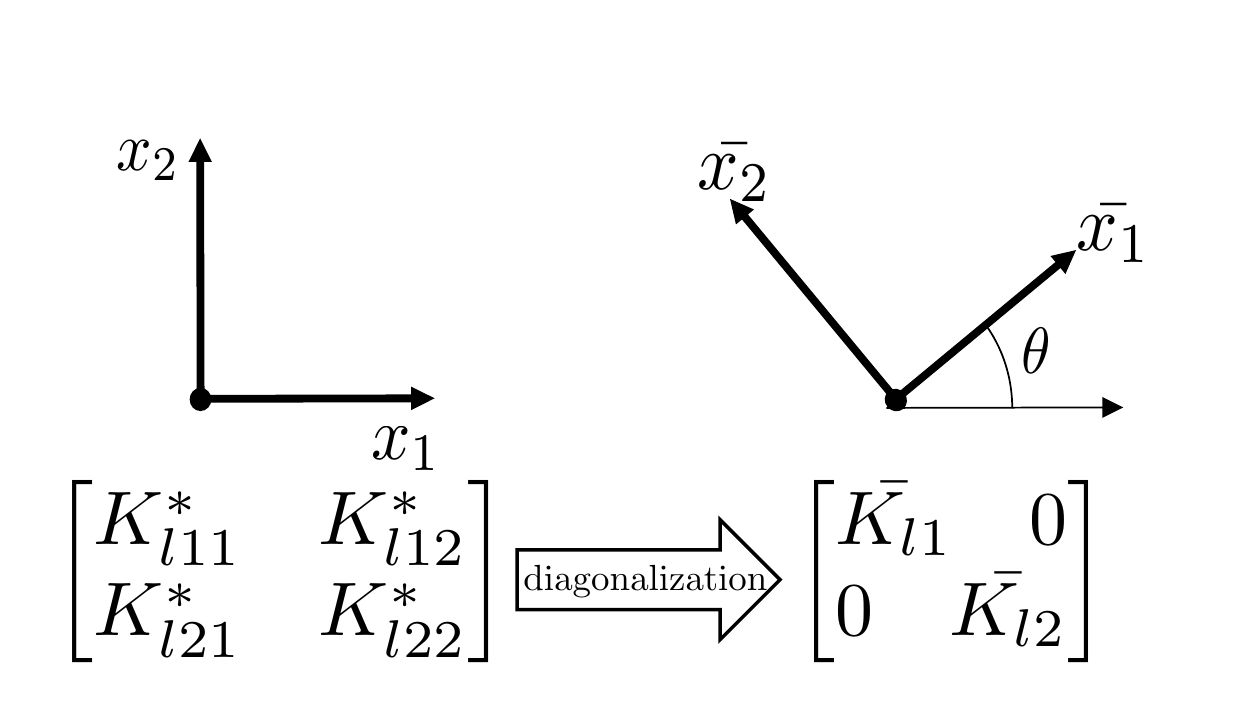}
	\caption{Diagonalization and rotation of the coordinate system.}
	\label{fig:diagonalization}
\end{figure}
Fig.~\ref{fig:initial struct}(e) shows the magnitude and direction of the heat flux. From this figure, we observed that the magnitude of the heat flux is large at $\Omega_C$. By minimizing the objective function $J_2$, we obtained a small magnitude of the heat flux in $\Omega_C$.
In Section~\ref{subsec:w=1}, we offer an optimization result with $w=1$, where $J_1$ is only minimized.
In Section~\ref{subsec:w=2}, an optimization case with $w=\frac{1}{2}$ is shown, where both $J_1$ and $J_2$ are considered in the optimization.
Furthermore, the validity of the optimized structure is discussed in Section~\ref{subsec:The validity of the optimization }.
\subsection{Optimization result with $w=1$}\label{subsec:w=1}
\begin{figure}[H]
	\centering
	\includegraphics[width=1\linewidth]{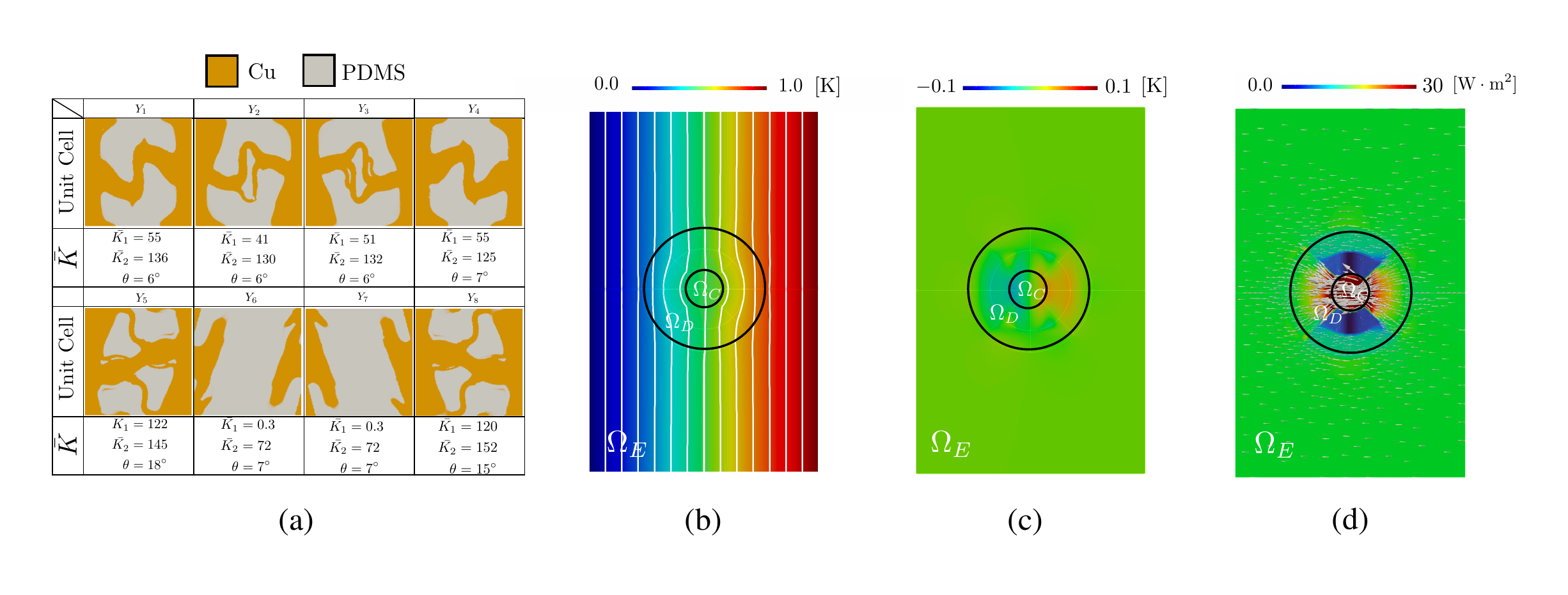}
	\caption{\textcolor{black}{(a) The unit cell structure of $Y_l$ with $1 \le l \le 8$ and the diagonalized thermal conductivity tensor $\bar{K}$. (b) The distribution of the temperature field $T$ obtained through homogenization. (c) The distribution of the temperature difference $T_\text{sub}$. (d) The magnitude and direction of the heat flux.} }
\label{fig:optimizationresultw1}
\end{figure}
\begin{figure}[H]
	\centering
	\includegraphics[width=0.45\linewidth]{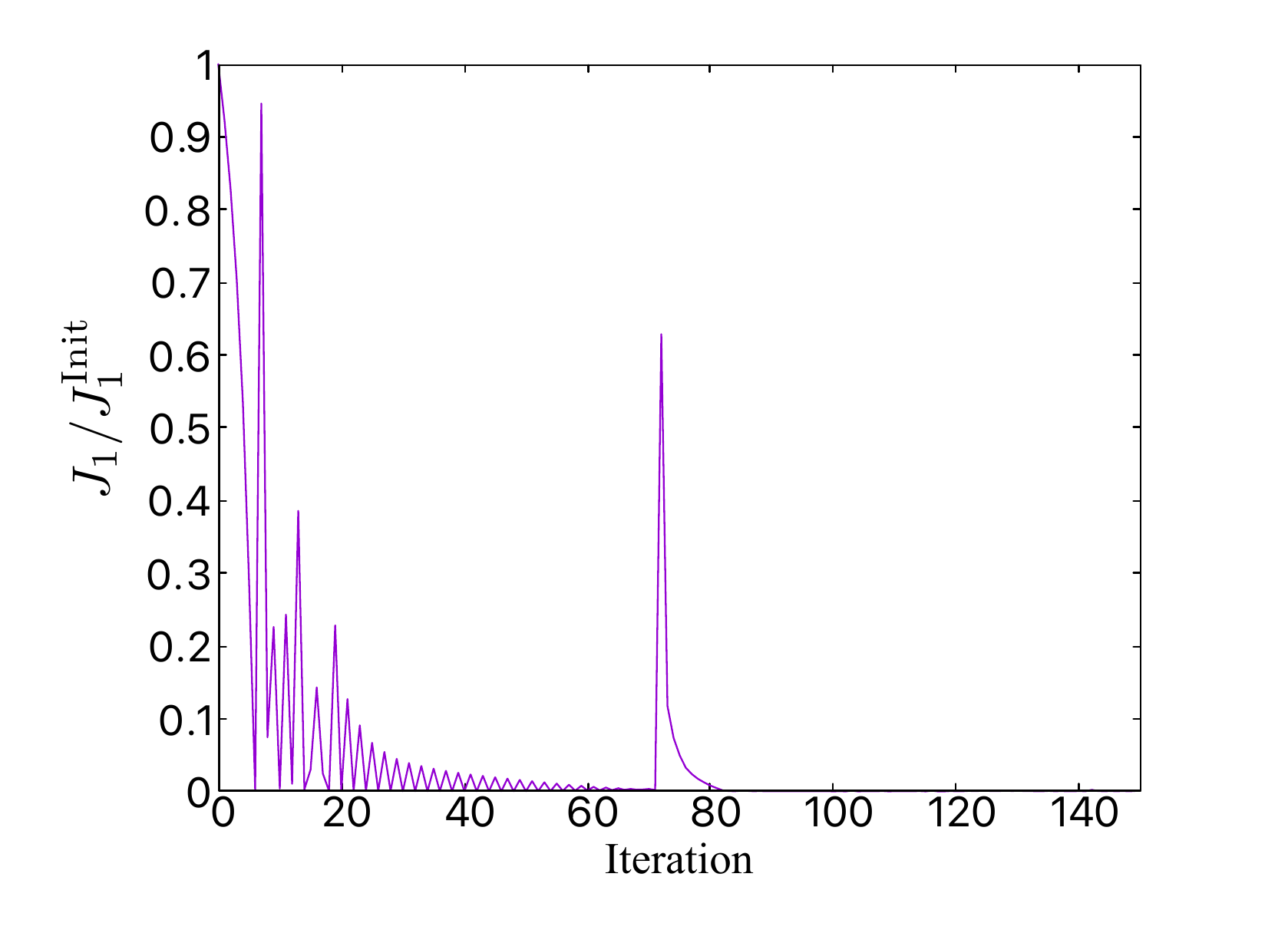}
	\caption{ \textcolor{black}{History of a nondimensional objective function $J_1/J_1^{\text{Init}}$.}}
	\label{fig:iter_obj_w1}
\end{figure}
When the weighting coefficient is $w=1$, the objective function to be minimized is $J=J_1$.
Fig.~\ref{fig:optimizationresultw1}(a) shows the optimized structure of each of the design domains $Y_l$ with $1 \le l \le 8$ and the diagonalized thermal conductivities tensor $\bar{K}$ with the rotation angle $\theta$. Fig.~\ref{fig:optimizationresultw1}(b, c) shows the temperature field obtained through homogenization and using the temperature difference $T_\text{sub}$, respectively.
From  Fig.~\ref{fig:optimizationresultw1}(b), the optimization result shows that $T_\text{sub}$ approaches $0$ in the evaluation domain $\Omega_E$; that is, the temperature field in the evaluation domain $\Omega_E$ approaches $T_\text{steel}$.
\textcolor{black}{Furthermore, a comparison of the objective function's values shows that the value of} 
\textcolor{black}{$J_{1}^{\text{Init}}=2.22\times10^{-2}\ \text{K}^2\text{m}^2$ in the initial structure is reduced to \textcolor{black}{$J_1=2.46\times10^{-6}\ \text{K}^2\text{m}^2$}.}
\textcolor{black}{The ratio
of the objective function's value of the optimized structure to that of the initial structure $ J_1/J_{1}^{\text{Init}}$ is $1.1\times10^{-4}$.}
\textcolor{black}{Fig.~\ref{fig:iter_obj_w1} shows the history of the objective function $ J_1/J_{1}^{\text{Init}}$. As explained in Section 4.2, we changed the value of the transition width $d$ used in the approximated characteristic function at the 70th iteration to stabilize the optimization calculation. Consequently, a drastic increase in the objective function there was observed. 
However, the objective function's value decreased after a few steps from the 70th iteration, and it finally converged to a sufficiently smaller value than the initial one.}
From Fig.~\ref{fig:optimizationresultw1}(a), the design domains $D_2,D_3,D_6$ and $D_7$ have smaller values of $\bar{K_1}$ than $\bar{K_2}$. Comparing these with those in  Fig.~\ref{fig:optimizationresultw1}(d), we can see that the heat flux is smaller in these domains. This result indicates that the heat flux is reduced in $D_2,D_3,D_6,$ and $D_7$ to achieve a thermal cloak only. 
\textcolor{black}{Note that the optimized structure strongly depends on the initial structure as is the case with a gradient-based topology optimization method, which finds a local solution rather than the global one. 
Therefore, we can obtain other types of optimized structures with different initial configurations. 
Even so, we could obtain one of the local solutions with good thermal cloaking performance because the value of the objective function with the optimized structure was sufficiently reduced compared to the initial one. We obtained complex microstructures due to the setting of $D_l~(l=1,...,8)$ and the design sensitivity shown in Eq. (15) and (16).
According to Eq. (15) and (16), 
microstructures are updated depending on the temperature gradient and adjoint fields in the macroscale. 
Consequently, we obtained each optimized microstructure in $Y_l$ for controlling the heat flux direction in the macroscale domain $D_l$, where the optimized microstructure is periodically arrayed.
Since the integration region $D_l$ in Eq. (15) and (16) has a large area in the current settings, and the direction of heat flux in $D_l$ is nonuniform, each microstructure in $D_l$ must control heat fluxes in various directions to realize thermal cloaking.
Therefore, simple structures represented by fiber-like microstructures could not be obtained as the optimized results. }

\textcolor{black}{
We also conducted topology optimization in \ref{sec:fuji} with the same geometry and boundary conditions as in the previous study to demonstrate the proposed method's effectiveness~\cite{hirasawaExperimentalDemonstrationThermal2022}.
}
\subsection{Optimization result with $w=\frac{1}{2}$}\label{subsec:w=2}
\begin{figure}[H]
	\centering
	\includegraphics[width=1\linewidth]{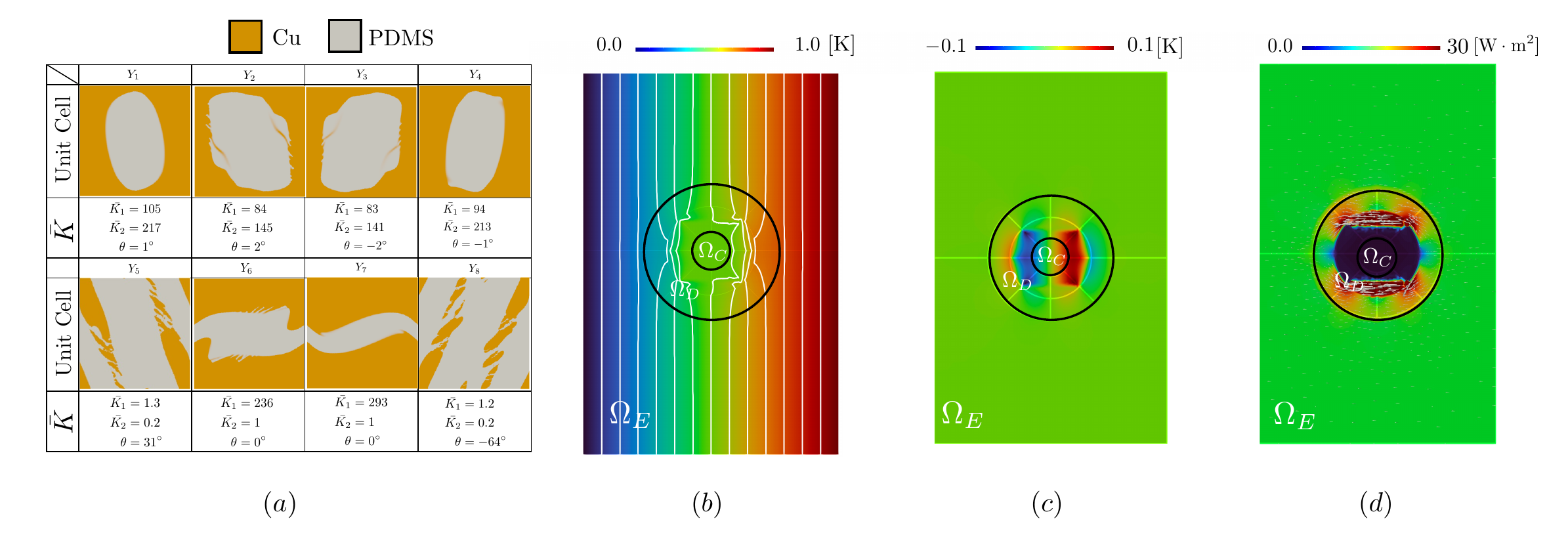}
	\caption{\textcolor{black}{(a) The unit-cell structure of $Y_l $ with $1 \le l \le 8$ and the diagonalized thermal conductivity tensor $\bar{K}$. (b) The distribution of the temperature field $T$ obtained through homogenization. (c) The distribution of the temperature difference $T_\text{sub}$. (d) The magnitude and direction of the heat flux. }}
	\label{fig:optimization result　w2}
\end{figure}
\begin{figure}[H]
	\centering
	\includegraphics[width=0.75\linewidth]{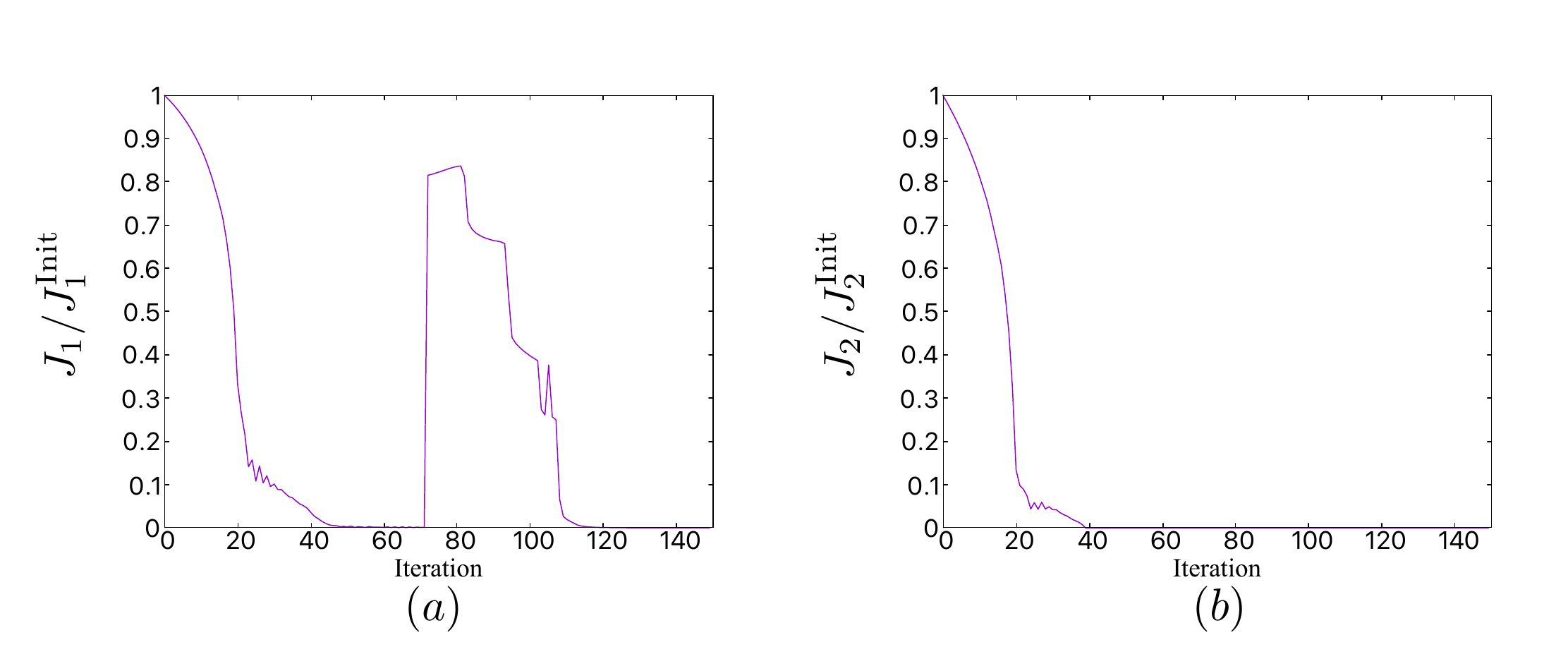}
	\caption{ \textcolor{black}{
 History of nondimensional objective functions. 
 (a) History of $J_1/J_1^{\text{Init}}$. 
 (b) History of $J_2/J_2^{\text{Init}}$. } }
	\label{}
\end{figure}
Next, the optimized structure is shown when the weight coefficient is $w=\frac{1}{2}$, that is, when the objective function to be minimized is $J=\frac{1}{2}J_1+\frac{1}{2}J_2$.
Fig.~\ref{fig:optimization result　w2}(a) shows the optimized structure of each of the design domains $Y_l$ with $1 \le l \le 8$ and the diagonalized thermal conductivities tensor $\bar{K}$ and the rotation angle $\theta$. 
Fig.~\ref{fig:optimization result　w2}(b, c) shows the temperature field obtained through homogenization and using the temperature difference $T_\text{sub}$, respectively. 
From Fig.~\ref{fig:optimization result　w2}(c), the optimization shows that $T_\text{sub}$ approaches $0$ in the evaluation domain $\Omega_E$; that is, the temperature field in the evaluation domain $\Omega_E$ approaches $T_\text{steel}$.
\textcolor{black}{The value of} 
\textcolor{black}{$J_1=4.22\times 10^{-6}\ \text{K}^2\text{m}^2$}
\textcolor{black}{is larger than that in the results shown in Section~\ref{subsec:w=1} because of the additional objective function $J_2$, but it has decreased compared with the initial structure.}
\textcolor{black}{The ratio of the objective function's value of the optimized structure to that of the initial structure $ J_1/J_{1}^{\text{Init}}$ is $1.9\times10^{-4}$.}
 \textcolor{black}{The value of} 
 \textcolor{black}{$J_2=2.47\times 10^{-9}\ \text{K}^2$ is also smaller than that of the initial structure, where the nondimensional objective function $J_2/J_{2}^{\text{Init}}$ is estimated as $1.72\times10^{-6}$.}
\textcolor{black}{
Fig.~9 (a) and (b) show the history of the objective function. 
As in the optimization result with $w=1$, a drastic increase in the value of $J_1/J_{1}^{\text{Init}}$ was observed at the 70th iteration due to the change in the value of $d$. 
However, both $J_1$ and $J_2$ decreased after a few steps from that iteration, and they finally converged to sufficiently smaller values than those of the initial structure.}
Compared with the result in Section~\ref{subsec:w=1}, Fig.~\ref{fig:optimization result　w2}(d) shows that the heat flux entering $\Omega_C$ is reduced by adding $J_2$ to the objective function.
\textcolor{black}{
Comparing the diagonalized thermal conductivity tensor with rotation angle $\theta$ in Fig.~8(a -- d), 
the diagonalized thermal conductivity tensor $\Bar{K_1}$ becomes smaller than $\Bar{K_2}$ in the domains $D_1$ and $D_4$. Therefore, the heat flux in these regions points to the $x_2$ direction. 
However, the heat flux in regions $D_6$ and $D_7$ flows only in the $x_1$ direction because 
$\Bar{K_1}$ is much larger than $\Bar{K_2}$.
Furthermore, the domains $D_5$ and $D_8$ are almost thermally insulated because the values of $\Bar{K_1}$ and $\Bar{K_2}$ approach zero in these regions. 
The macroscopic thermal conductivity tensor distribution successfully
reduced the heat flux entering $\Omega_C$.
}
\section{Validity of the optimized structures} \label{subsec:The validity of the optimization }
In Sections~\ref{subsec:w=1} and \ref{subsec:w=2}, the homogenization method was used to find the optimized structure. As explained in Section~\ref{sec: homogenization}, this method assumes that infinitesimally small unit cells are arranged in the design domain $D_l$ with $1 \le l \le8$. However, in the actual design, creating unit cells with infinitesimally small sizes as assumed in the homogenization method is impossible. Therefore, we investigated the validity of the optimized structure by examining the temperature field when the unit cells $Y_l$ of finite sizes are placed in the design domain $D_l$ with $1 \le l \le8$. 
\subsection{Arrangement of finite size unit cells}\label{subsec:the arrangment of finite size}
\begin{figure}[H]
	\centering
	\includegraphics[scale=0.8]{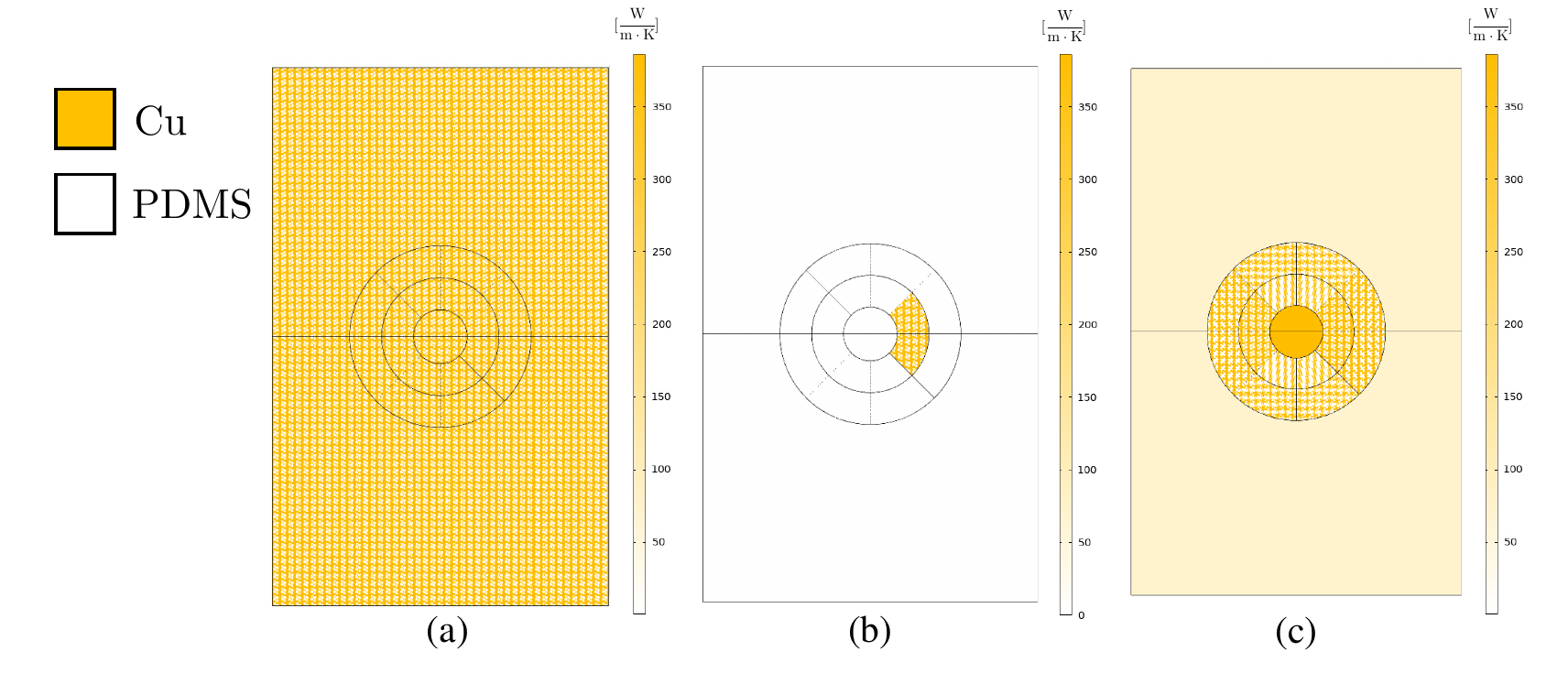}
	\caption{\textcolor{black}{Laid out finite unit cell. The legends represent the value of the thermal conductivity tensor's value. (a) The unit cell $Y_l$ obtained in Section~\ref{sec: numerical examples} is laid out in all domains. (b) The unit cells $Y_l$ are lined up only in $D_l$. (c) Thermal conductivity when the unit cells of finite sizes are arranged in all $D_1 -- D_8$. }}
	\label{fig:finitecell}
\end{figure}
The unit cells $Y_1$ obtained by optimization are laid out, as shown in Fig.~\ref{fig:finitecell}(a).
We first map the material distribution in $Y_l$ to the whole macroscale domain, as shown in Fig.~\ref{fig:finitecell}(a) with $l = 1$, to assign the material distribution defined in each unit cell $Y_l$ with $1\le l \le 8$ in the macroscale domain $D_l$.
\textcolor{black}{The size of a unit cell is $\epsilon_0=\frac{1}{9} \ \text{m} $. }
Then, we only focused on the mapped material distribution in $D_l$, as shown in Fig.~\ref{fig:finitecell}(b).
These procedures were repeated for $1\le l \le 8$, and we assigned steel and copper to $\Omega_E$ and $\Omega_C$, respectively.
Finally, the material distribution based on the microstructures was obtained, as shown in Fig.~\ref{fig:finitecell}(c).
According to this material distribution, the temperature field is obtained by solving the heat conduction problem using FEM without the homogenization method. 
When the unit cells were the initial structure shown in Fig.~\ref{fig:initial struct} with finite size, the values of the objective function were evaluated as  
$J_1^{\text{Init}}=2.0\times10^{-2}\ \text{K}^2\text{m}^2$, $J_2^{\text{Init}}=1.5\times10^{-3}  \ \text{K}^2$.
Next, using this method, the values of the objective function were examined for the optimized structures in Sections~\ref{subsec:w=1} and \ref{subsec:w=2}. 
\begin{figure}[H]
	\centering
	\includegraphics[width=1\linewidth]{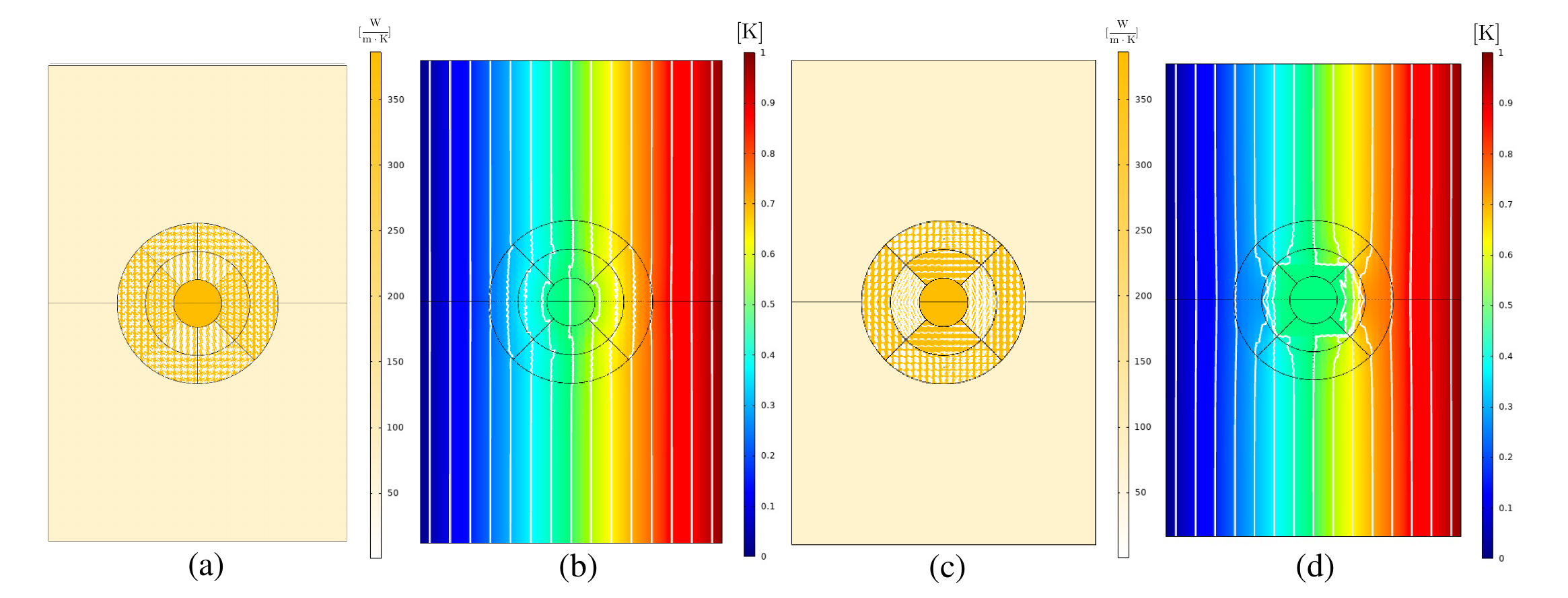}
	\caption{\textcolor{black}{(a) Finite-size optimized unit-cell structures of Section~\ref{subsec:w=1} are arranged. (b) The temperature field distribution of (a). (c) Finite-size optimized unit-cell structures of Section~\ref{subsec:w=2} are arranged. (d) The temperature field distribution of (c). }} 
	\label{fig:w1w2}
\end{figure}
Fig.~\ref{fig:w1w2}(a) shows the distribution of the thermal conductivity when the optimized structures of Section~\ref{subsec:w=1} are lined up. 
Fig.~\ref{fig:w1w2}(b) shows the temperature field obtained by solving Eq.~(\ref{eq: g.v.e.q}) using the thermal conductivity distribution in Fig.~\ref{fig:w1w2}(a). 
\textcolor{black}{
The values of the objective functions $J_1$ and $J_2$ were 
$J_1=4.6\times 10^{-6}\ \text{K}^2\text{m}^2$ and $J_2=2.3\times 10^{-3}\ \text{K}^2$, 
respectively. 
Thus, $ J_1/J_{1}^{\text{Init}}$ is estimated as is $2.3\times10^{-4}$.}
This certainly shows that a thermal cloak has been achieved. 
Fig.~\ref{fig:w1w2}(c) shows the distribution of the thermal conductivity when the optimized structures of Section~\ref{subsec:w=2} are lined up. 
Fig.~\ref{fig:w1w2}(d) shows the temperature field obtained by solving Eq.~(\ref{eq: g.v.e.q}) for Fig.~\ref{fig:w1w2}(c).
\textcolor{black}{The values of the objective functions $J_1$ and $J_2$ were 
$J_1=3.7\times 10^{-5}\ \text{K}^2\text{m}^2$ and $J_2=4.6\times10^{-9}\ \text{K}^2$, respectively.}
\textcolor{black}{Thus, $J_1/J_{1}^{\text{Init}}$ and  $J_2/J_{2}^{\text{Init}}$ are estimated as $1.9\times10^{-3}$ and $3.0\times10^{-6}$, respectively.}
\textcolor{black}{The temperature field distribution in Fig.~\ref{fig:w1w2}(b) and (d) differ from those in Fig.~\ref{fig:optimizationresultw1}(b) and Fig.~\ref{fig:optimization result　w2}(b), and the value of the objective functions $J_1$ and $J_2$ in Section~6 differ from those in Section~5. 
This result is because here we focus on the cloaking structure comprising optimized unit cells with finite size. The temperature distribution in Fig.~\ref{fig:w1w2}(b) and (d) was obtained by solving the heat conduction equation without the homogenization method, which is different from Section $5$. However, comparing the objective function's values for the initial structure, a thermal cloak and no heat flux in $\Omega_C$ have been achieved even if the unit cell has a finite size.
}
As we can see from Fig.~\ref{fig:w1w2}(b) and (d), the structural boundaries around the boundaries of the design region $D_l$ with $1\le l \le 8$ are not continuous in some regions or are overlapped in some regions. If these effects on the distribution of the temperature fields are considered during the optimization process, the value of the objective function could change and the optimized structure would differ from the above results in Sections~\ref{subsec:w=1} and \ref{subsec:w=2}. Therefore, there is room for discussion, for example, using high order homogenization instead of the standard homogenization used in this research.
However, the optimized structures obtained in Section~\ref{sec: numerical examples} are generally appropriate because the value of the objective functions are smaller than those of the initial configuration even if the cell is of finite size without assuming an infinitely small unit cell as in the homogenization method.
\subsection{Robustness of the optimized structures}
\begin{figure}[H]
	\centering
	\includegraphics[scale=0.55]{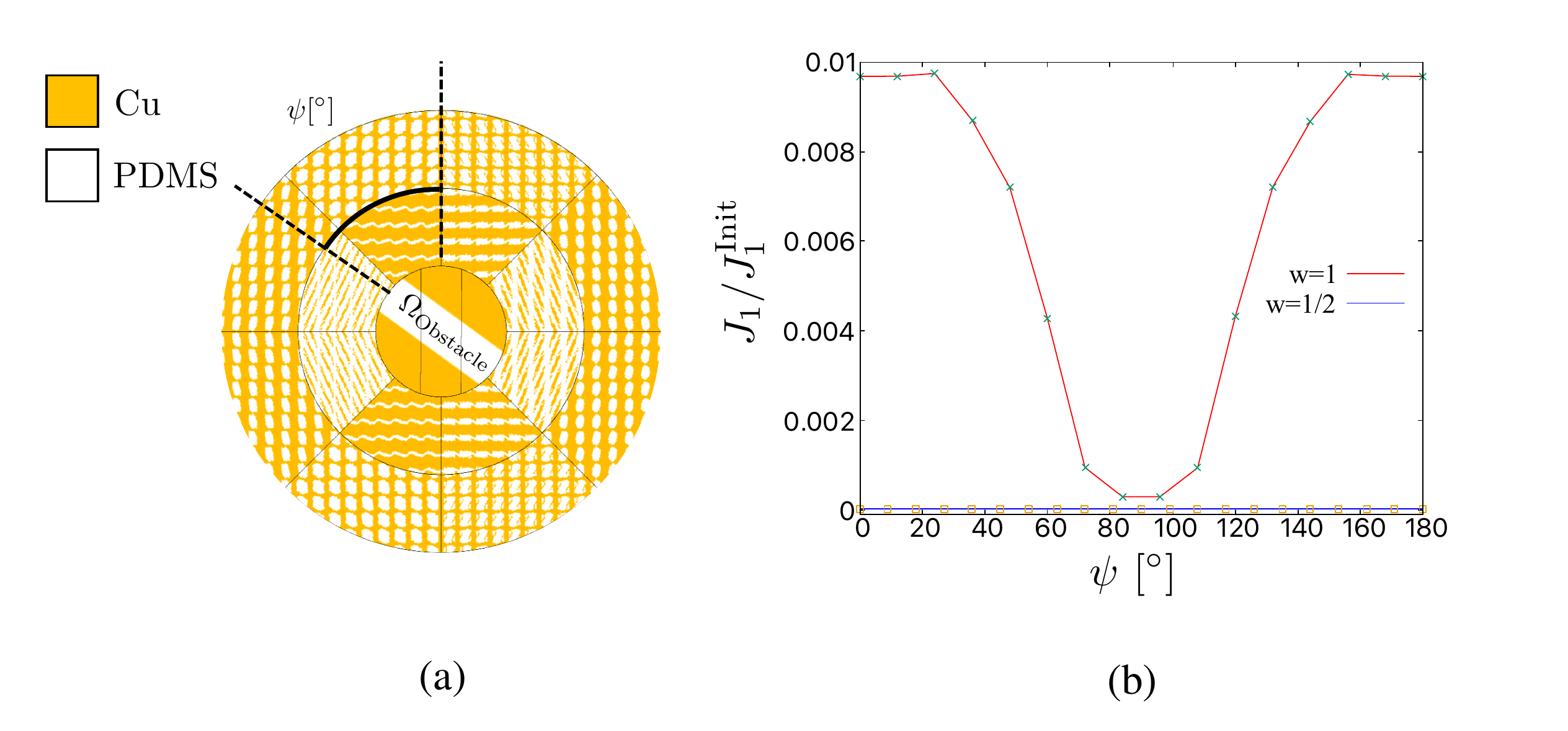}
	\caption{\textcolor{black}{(a) Finite-size optimized unit-cell structures are arranged and an obstacle made of PDMS is placed in $\Omega_C$ at different $\psi ^\circ$. (As an example, the optimized structures in Section 5.3 are shown in this image.) (b) Response of the objective function $J_1/J_1^{\text{Init}}$ with respect for $\psi^\circ$.
}}
	\label{fig:Robust}
\end{figure}
\textcolor{black}{
The optimized structures in Sections~\ref{subsec:w=1} and \ref{subsec:w=2} are laid out as in Section~6.1, and we placed an object made of PDMS in $\Omega_C$ as shown in Fig.~\ref{fig:Robust}(a).}
\textcolor{black}{Based on this setting, we} examined the change in the value of the objective function $J_1$ as a function of $\psi$ to confirm the meaning of adding the objective function $J_2$. The relationship between $\psi$ and $J_1$ is shown in Fig.~\ref{fig:Robust}(b). 
When the objective function is $J=J_1$, the cloaking performance varies depending on the angle $\psi$ of the obstacle, as can be seen from the figure. 
\textcolor{black}{Furthermore, the value of the objective function $J_1$ with $w=1$ is also larger than that in Section~6.1 because the optimized structure in Section~\ref{subsec:w=1} realizes thermal cloaking by focusing the heat flux into $\Omega_C$. Since there is an obstacle in $\Omega_C$ to block the heat flux, the cloaking performance became worse than in Section~6.1.}
However, by adding $J_2$ to the objective function to be minimized, 
\textcolor{black}{
we can confirm that the value of $J_1/J_1^{\text{Init}}$ is close to $0$ even when $\psi$ is varied, in other words, 
}
a thermal cloak is robust to the object placed at $\Omega_C$. This is because the heat flux at $\Omega_C$ is small; therefore, no disturbance of the heat flux due to the placement of an object at $\Omega_C$ occurs.
From the above, it can be seen that minimizing $J_1$ and $J_2$ simultaneously as the objective function enables a thermal cloak no matter what obstacle is placed.
\section{Conclusion} \label{sec: conclusion}
This study proposed a level set-based topology optimization method for the design of a thermal cloak \textcolor{black}{based on artificially designed composite materials}. The results obtained by this study are as follows:
\begin{enumerate}
	\item On the basis of previous studies, the homogenization method was introduced for the heat conduction problem, and the optimization problem was formulated such that the unit cell structures composing the \textcolor{black}{composite materials} were set as the design variables.  This enabled multiscale optimization for the thermal cloak design problem, where the objective function defined using the macroscale temperature field is minimized.
	
	\item A level set-based topology optimization method was introduced to solve the optimization problem, and sensitivity analysis based on the concept of topological derivatives was performed.
	\item According to the formulation of the optimization problem, the optimization algorithm was constructed.
	\item The proposed method provided optimized unit cell structures that achieved a thermal cloak and no heat flux simultaneously. In other words, by minimizing the objective functions formulated in Section~\ref{sec:formulation}, optimized structures that simultaneously achieve two phenomena were obtained: the temperature field outside the cloak approaches the reference temperature field while heat flux is reduced inside the cloak.
	\item The validity of the proposed method was demonstrated by solving the heat conduction problem for the entire domain using the optimized unit cells of finite sizes.
\end{enumerate}

In future works, this method will be applied to the three-dimensional thermal cloak design problem, which will allow us to consider systems that are closer to real-world problems. Because the division of microstructures within macrostructures was arbitrary, the number of unit cells can be increased, or the division can be done in consideration of the magnitude and direction of the heat flux, which is expected to improve the thermal cloak and the no heat flux performance.
\appendix
\section{Sensitivity analysis details}\label{sec:Appendix sensitivity analysis}
This section describes how we obtained the design sensitivity of the response of the homogenization coefficient. In other words, we obtained the derivative of the objective function described in Section~\ref{subsec:sensitivity analysis} with respect to the thermal conductivity tensor $K^*$.
Corresponding to the optimization problem, the Lagragian $L$ is defined as follows:
\begin{align}
	&L(K_1^*,...,K_8^*, \hat{T}, \hat{v^k_H}, \hat{\lambda_a}, \hat{\lambda_b}, \hat{\lambda_c})\nonumber\\
	&=J_k(\hat{T})+\sum_{l=1}^8\int_{D_l}\nabla_x \cdot(K_l^*\nabla_x \hat{T})\hat{v^k_H}d\Omega+\int_{\Omega_\text{ext}}\nabla_x \cdot({K}^{}\nabla_x \hat{T})\hat{v^k_H}d\Omega  \nonumber\\
	&-\int_{\Gamma_a}(\hat{T}_H-u_a)\hat{\lambda}^ad\Gamma-\int_{\Gamma_b}(\hat{T}_H-u_b)\hat{\lambda}^bd\lambda-\int_{\Gamma_N}\bm{n}\cdot({K}\nabla_x \hat{T})\hat{\lambda}^cd\Gamma,
\end{align}
where  $K_l^*$ with $1\le l \le 8$ is the homogenized thermal conductivity evaluated using  Eq.~(\ref{eq:homogezied coefficient}), $\hat{v^k_H}$, $\hat{\lambda_a}, \hat{\lambda_b},$ and $  \hat{\lambda_c}$ are Lagrange multipliers, and $\hat{T}$ is a variable corresponding to the macroscopic temperature field.
$J_k(\hat{T})$ is the objective function, generalized from the two types of objective functions $J_1$, and  $J_2$ defined in Section~\ref{sec:formulation} and given in the following form:
\begin{align}
J_k(\hat{T})=\int_{\Omega_\text{meas}}j_k(\hat{T})d\Omega \label{eq:JT}\ (k=1, 2),
\end{align}
where $\Omega_\text{meas}$ is the domain that evaluates the value of the objective function using the temperature field, specifically $\Omega_E$ for $J_1$ and $\Omega_C$ for $J_2$. 
$\Omega_\text{ext}$ represents the region excluding $\Omega_\text{meas}$ and $\Omega_D$ from the entire computational domain.
At the stationary point of the Lagrangian, denoted by $opt$, the following optimality conditions (\ref{eq:LT})--(\ref{eq:Lc}) hold:
\begin{eqnarray}
	\left<  \left. \frac{\partial L}{\partial \hat{T}}, \delta \hat{T}\right>\right|_{opt}=0 \label{eq:LT},\\
	\left< \left. \frac{\partial L}{\partial \hat{v^k_H}}, \delta \hat{v^k_H}\right>\right|_{opt}=0 \label{eq:Lv},\\
	\left<  \left.\frac{\partial L}{\partial \hat{\lambda_a}}, \delta \hat{\lambda_a}\right>\right|_{opt}=0 \label{eq:La},\\
	\left<  \left.\frac{\partial L}{\partial \hat{\lambda_b}}, \delta \hat{\lambda_b}\right>\right|_{opt}=0 \label{eq:Lb},\\
	\left<  \left.\frac{\partial L}{\partial \hat{\lambda_c}}, \delta \hat{\lambda_c}\right>\right|_{opt}=0 \label{eq:Lc}.\\
	\nonumber
\end{eqnarray}
The optimality conditions (\ref{eq:Lv})--(\ref{eq:Lc}) are satisfied if $\hat{T}=T$ at the stationary point of the Lagrangian.
Then, the optimality condition in (\ref{eq:LT}) is expressed as follows:
\begin{align}
	\left< \left. \frac{\partial L}{\partial \hat{T}}, \delta T\right >\right|_{opt}
	&=\left< \left. \frac{\partial J}{\partial \hat{T}}, \delta T \right>\right|_{opt}+\sum_{l=1}^8 \int_{D_l}\delta T(\nabla_x(K_l^*\nabla v^k_H))d\Omega + \int_{\Omega_\text{ext}}\delta T(\nabla_x(K\nabla v^k_H))d\Omega \nonumber\\
	&-\int_{\Gamma_a}\delta T(\bm{n}\cdot(K\nabla_x v^k_H)+\lambda_a)d\Gamma \nonumber+\int_{\Gamma_N}\bm{n}\cdot(K\nabla_x v^k_H)\delta T d\Gamma \\
	&+\int_{\Gamma/\Gamma_N}\bm{n}\cdot(K\nabla_x \delta T)v^k_Hd\Gamma\nonumber \\
	&-\int_{\Gamma_N}\bm{n}\cdot(K\nabla_x v^k_H)\delta T d\Gamma -\int_{\Gamma_N}\bm{n}\cdot(K\nabla_x \delta T)(\lambda_c-v^k_H)d\Gamma=0.
	\label{eq:a.7}
\end{align}
This equation holds if the following adjoint equation (\ref{eq:Adjoint eq}) holds: 
\begin{eqnarray}
	\left\{
	\begin{array}{ll}
		-\nabla_x(K\nabla_x v^k_H)&=0\qquad \text{in} \qquad \Omega_\text{ext}\backslash \overline{\Omega_\text{meas}}, \\
		-\nabla_x(K\nabla_x v^k_H)&=  \frac{\partial j_k(T)}{\partial T}\qquad \text{in} \qquad \Omega_\text{meas}, \\
		-\nabla_x(K_l^*\nabla_x v^k_H)&=0\qquad \text{in} \qquad D_l \ (1 \le l \le 8),\\
		v_H &=0 \qquad \text{on} \qquad \Gamma_a, \Gamma_b,\\
		\bm{n}\cdot(K^*\nabla_x v^k_H) &=0 \qquad \text{on}\qquad\Gamma_N.
	\end{array}
	\right.
	\label{eq:Adjoint eq}
\end{eqnarray}
Furthermore, the following relationships of the Lagrange multipliers should be satisfied at the stationary point to fulfill Eq.~(\ref{eq:LT}):
\begin{eqnarray}
	\left\{
	\begin{array}{ll}
		\lambda_a&=\lambda_b=-\bm{n}\cdot(K^*\nabla_x v^k_H) \qquad \text{on} \qquad \Gamma_a, \Gamma_b, \\
		\lambda_c&=v^k_H\qquad \text{on} \qquad \Gamma_N.
	\end{array}
	\right.
	\label{eq: ablambda}
\end{eqnarray}
Due to the definition of $L$, we get  $L(K^*, T, \hat{v_H^k})=J(T)$. Therefore, the derivative of the objective function with respect to the thermal conductivity tensor  is as follows:
\begin{align}
	\frac{\partial J_k}{\partial  K_{lij}^*}
	&=\frac{\partial L(K^\ast, T, \hat{v^k_H})}{\partial K_{lij}^*} +\left<  \frac{\partial L(K^\ast, T, \hat{v^k_H})}{\partial \hat{T}}, \delta T^{'}(K_{lij}^*)\right>.
		\label{eq: derivative of J wrt Klij}
\end{align}
Therefore, the design sensitivity $\cfrac{\partial J_k}{\partial K^*_{lij}}$ is finally obtained using the optimality condition in Eq.~(\ref{eq:LT}), as follows.
\begin{align}
	\cfrac{\partial J_k}{\partial K^*_{lij}}=-\int_{D_l} \frac{\partial T}{\partial x_i} \frac{\partial v^k_H}{\partial x_j} d\Omega\ (k=1, 2).
\end{align}

\textcolor{black}{
\section{Thermal cloak optimization based on the setting in \cite{hirasawaExperimentalDemonstrationThermal2022} }\label{sec:fuji}
We conducted topology optimization for a thermal cloak with the same geometry and boundary condition settings as in the previous study~\cite{hirasawaExperimentalDemonstrationThermal2022} to demonstrate the superiority of our proposed method, which combines the homogenization method and topology optimization,  
In \cite{hirasawaExperimentalDemonstrationThermal2022}, the homogenization method was not incorporated with an optimization method, and topology optimization for thermal cloaks was conducted only on the macroscale. 
The proposed method's optimized result is compared with the previous one, especially for the cloaking performance. 
}

\begin{figure}[H]
	\centering
	\includegraphics[width=1\linewidth]{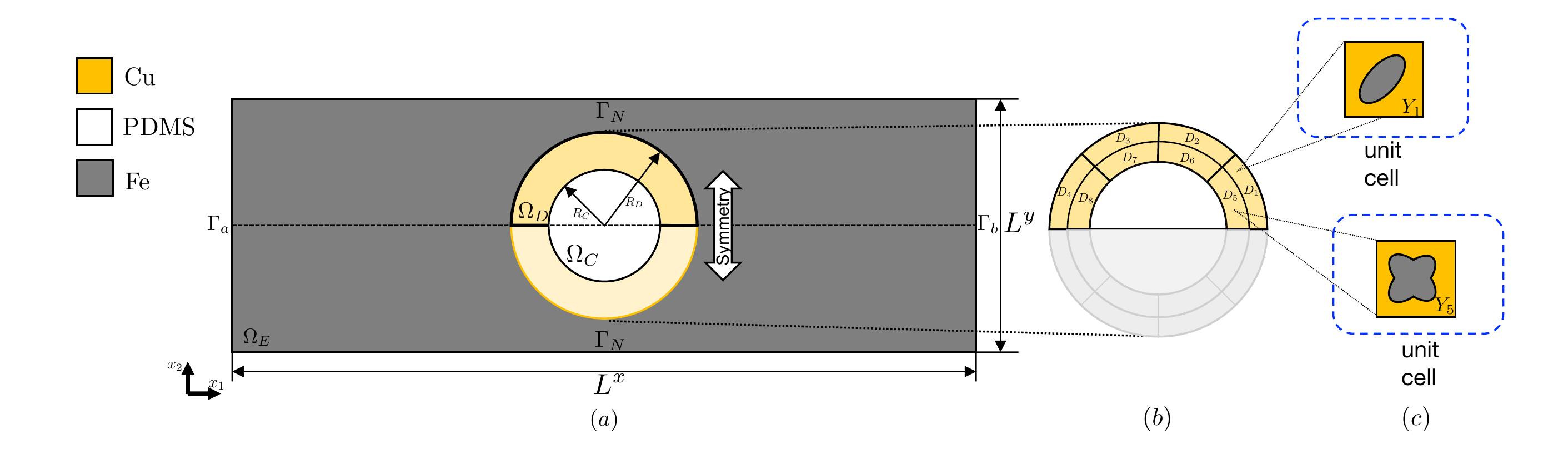}
	\caption{
 (a) Geometry settings of the thermal cloak design problem.  $\Omega_E$ is the evaluation domain, $\Omega_D$ is the domain containing the optimized microstructures, and $\Omega_C$ is a domain filled with PDMS. The structure's characteristic lengths are $R_D=5\ \text{m}$,  $L^x=20\ \text{m}$,  $L^y=\frac{40}{3}\ \text{m}$, $R_c=3\ \text{m}$.   (b) The magnified view around $\Omega_D$ divided into eight regions, $D_l$ with $1 \le l \le 8$. (c) A unit cell $Y_l$ with $ 1\le l \le 8$ with the infinitesimal size is periodically laid out in $D_l$.}
	\label{fig:fujiDesign_model}
\end{figure}
The topology optimization problem setting for the thermal cloak is presented in Fig.~\ref{fig:fujiDesign_model}. 
In the macroscale, the computational domain comprised $\Omega_E$, $\Omega_D$, and $\Omega_C$ in Fig.~\ref{fig:fujiDesign_model}(a).
$\Omega_E$ is the evaluation domain for the thermal cloak filled with steel.
$\Omega_D$ is the domain where the microstructures are placed, and it is divided into several design domains $D_l$ with $1\le l \le 8$ as shown in Fig.~\ref{fig:fujiDesign_model}(b).
As shown in Fig.~\ref{fig:fujiDesign_model}(c), a microstructure made of copper ($Y_l^c$) and steel ($Y_l^s$) defined in the unit cell $Y_l$ is assigned to the domain $D_l$ with $1\le l \le 8$. 
$\Omega_C$ is an obstacle for thermal cloaking and is assumed to be made of PDMS. The boundary condition on $\Gamma_N$ is the adiabatic condition, whereas those on $\Gamma_a$ and $\Gamma_b$ are the temperatures fixed to specific values (low temperature $T_\text{low}=0$ and high temperature $T_\text{high}=1$, respectively).
\textcolor{black}{
In this example, the objective function was defined as follows:
\begin{equation}
J = \cfrac{\int_{\Omega_{E}}(T - T_{\text{steel}})^2}{\int_{\Omega_{E}}(T_{\text{PDMS}} - T_{\text{steel}}) ^2},
\end{equation}
where  $T_{\text{steel}}$ is the reference temperature field when an entire domain is made of steel, $T_{\text{PDMS}}$ is the temperature field when the domain of $\Omega_D$ is made of PDMS. 
}
\begin{figure}[H]
	\centering
	\includegraphics[width=0.8\linewidth]{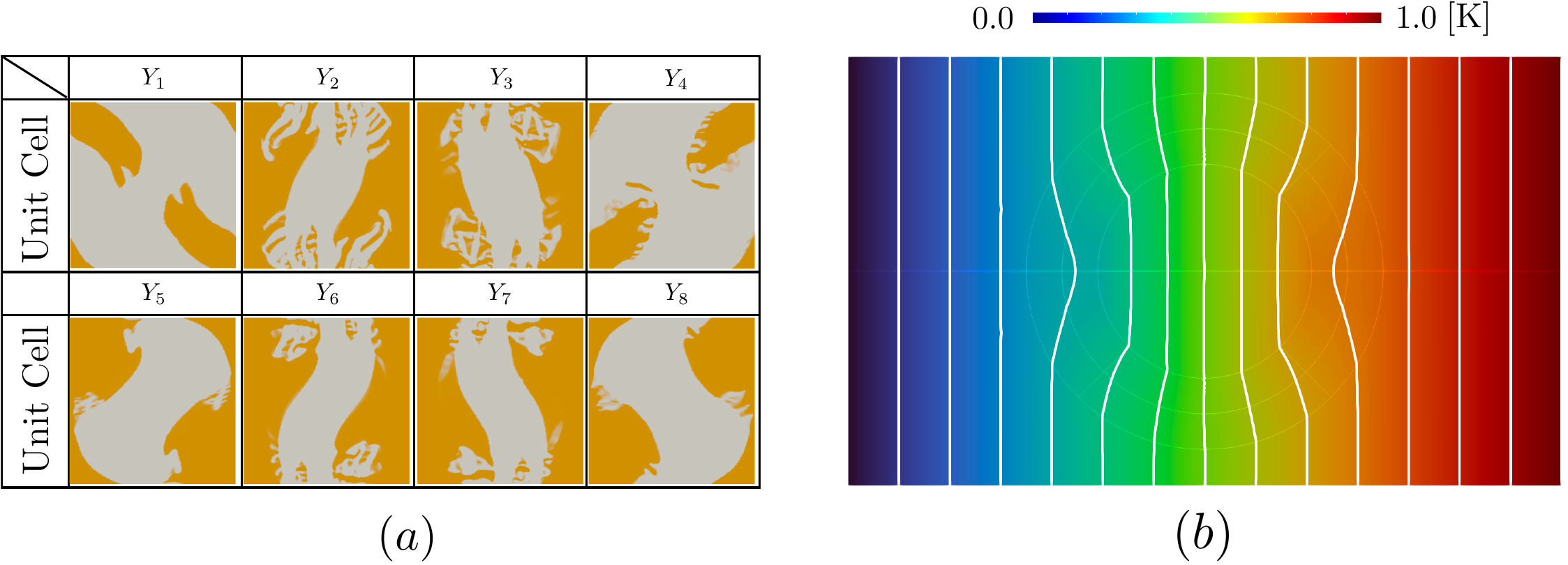}
	\caption{(a) Unit-cell structure of $Y_l $ with $1 \le l \le 8$. (b) Distribution of the temperature field $T$ obtained through homogenization.}
	\label{fig:fujiopt_result}
\end{figure}
\textcolor{black}{
Fig.~\ref{fig:fujiopt_result} shows the optimization results. 
Although the optimized structure is more complex than the previous one in \cite{hirasawaExperimentalDemonstrationThermal2022}, the optimized cloak in Fig.~\ref{fig:fujiopt_result}(a) exhibited better cloaking performance. 
The temperature distribution in Fig.~\ref{fig:fujiopt_result}(b) shows almost no temperature perturbation in the evaluation domain. 
The value of the objective function $J$ for the structure in Fig.~\ref{fig:fujiopt_result}(a) is $6.33\times 10^{-6}$, smaller than the objective function's value in ~\cite{hirasawaExperimentalDemonstrationThermal2022}, {$1.59\times 10^{-4}$}.
This is because the optimized results in the proposed method are based on the anisotropic macroscopic thermal conductivity realized by the microstructures. The previous result in \cite{hirasawaExperimentalDemonstrationThermal2022} was only based on the isotropic thermal conductivity of the constitutive materials. 
Therefore, the proposed method is beneficial to realize a thermal cloak with better performance compared to the previous method. 
}


\clearpage
\bibliographystyle{elsarticle-num}

\bibliography{Manuscript}

\end{document}